\documentclass[twocolumn,preprintnumbers,superscriptaddress,amsmath,amssymb,prb]{revtex4-1}

\usepackage{graphicx}
\usepackage{dcolumn}
\usepackage{subfigure}
\usepackage{bm}
\usepackage[utf8]{inputenc}
\usepackage{comment}
\usepackage[unicode=true, bookmarks=false, breaklinks=false, pdfborder={0 0 1},colorlinks=false]{hyperref}
\usepackage{breakurl}
\usepackage{url}
\usepackage{natbib}
\usepackage{multirow}
\usepackage{float}

\usepackage{wrapfig}

\usepackage{xcolor}

\newcommand{\unit}[2][1]{#1~\mathrm{#2}}

\begin{document}

\title{Tunable topological magnetism in superlattices of nonmagnetic B20 systems}

\newcommand{\Uppsala}{Department of Physics and Astronomy, Uppsala University, Box 516, SE-75120 Uppsala, Sweden}
\newcommand{\KTH}{Department of Applied Physics, School of Engineering Sciences, KTH Royal Institute of Technology, AlbaNova University Center, SE-10691 Stockholm, Sweden}
\newcommand{\WISEUppsala}{Wallenberg Initiative Materials Science for Sustainability, Uppsala University,
75121 Uppsala, Sweden}
\newcommand{\WISEKTH}{Wallenberg Initiative Materials Science for Sustainability (WISE), KTH Royal Institute of Technology, SE-10044 Stockholm, Sweden}
\newcommand{\SeRC}{SeRC (Swedish e-Science Research Center), KTH Royal Institute of Technology, SE-10044 Stockholm, Sweden}

\author{Vladislav Borisov}
    \affiliation{\Uppsala}
    \email[Corresponding author:\ ]{vladislav.borisov@physics.uu.se}

    \author{Anna Delin}
    \affiliation{\KTH}
    \affiliation{\WISEKTH}
    \affiliation{\SeRC}

\author{Olle Eriksson}
    \affiliation{\Uppsala}
    \affiliation{\WISEUppsala}
    
\date{\today}

\begin{abstract}
We predict topological magnetic properties of B20 systems, that are organized in atomically thin multilayers. In particular we focus on FeSi/CoSi and FeSi/FeGe superlattices with different number of layers and interface structure. We demonstrate that absence of long range magnetic order, previously observed in bulk FeSi and CoSi, is broken near the FeSi/CoSi interface, where a magnetic state with non-trivial topology appears. Using electronic structure calculations in combination with the magnetic force theorem, we calculate the Heisenberg and Dzyaloshinskii-Moriya (DM) interactions in these systems. With this information, we perform atomistic spin dynamics simulations at finite temperature and applied magnetic field for large supercells with up to $2\cdot10^6$ spins to capture the complexity of non-collinear textures induced by the DM interaction. The spin dynamics simulations predict the formation of antiskyrmions in a [001]-oriented FeSi/CoSi multilayer, intermediate skyrmions in a [111]-oriented FeSi/CoSi system and Bloch skyrmions in the FeSi/FeGe (001) system. The size of different types of skyrmions is found to vary between $\unit[7]{nm}$ and $\unit[37]{nm}$. The varying topological magnetic texture in these systems can be attributed to the complex asymmetric structure of the DM micromagnetic matrix, which is different from previously known topological magnets. Furthermore, through structural engineering, we demonstrate that both FM and AFM skyrmions can be stabilized, where the latter are especially appealing for applications due to the zero skyrmion Hall effect. The proposed B20 multilayers show potential for further exploration and call for experimental confirmation.
\end{abstract}

\maketitle

\begin{figure}
\vspace{10pt}
{\centering
\includegraphics[width=0.99\columnwidth]{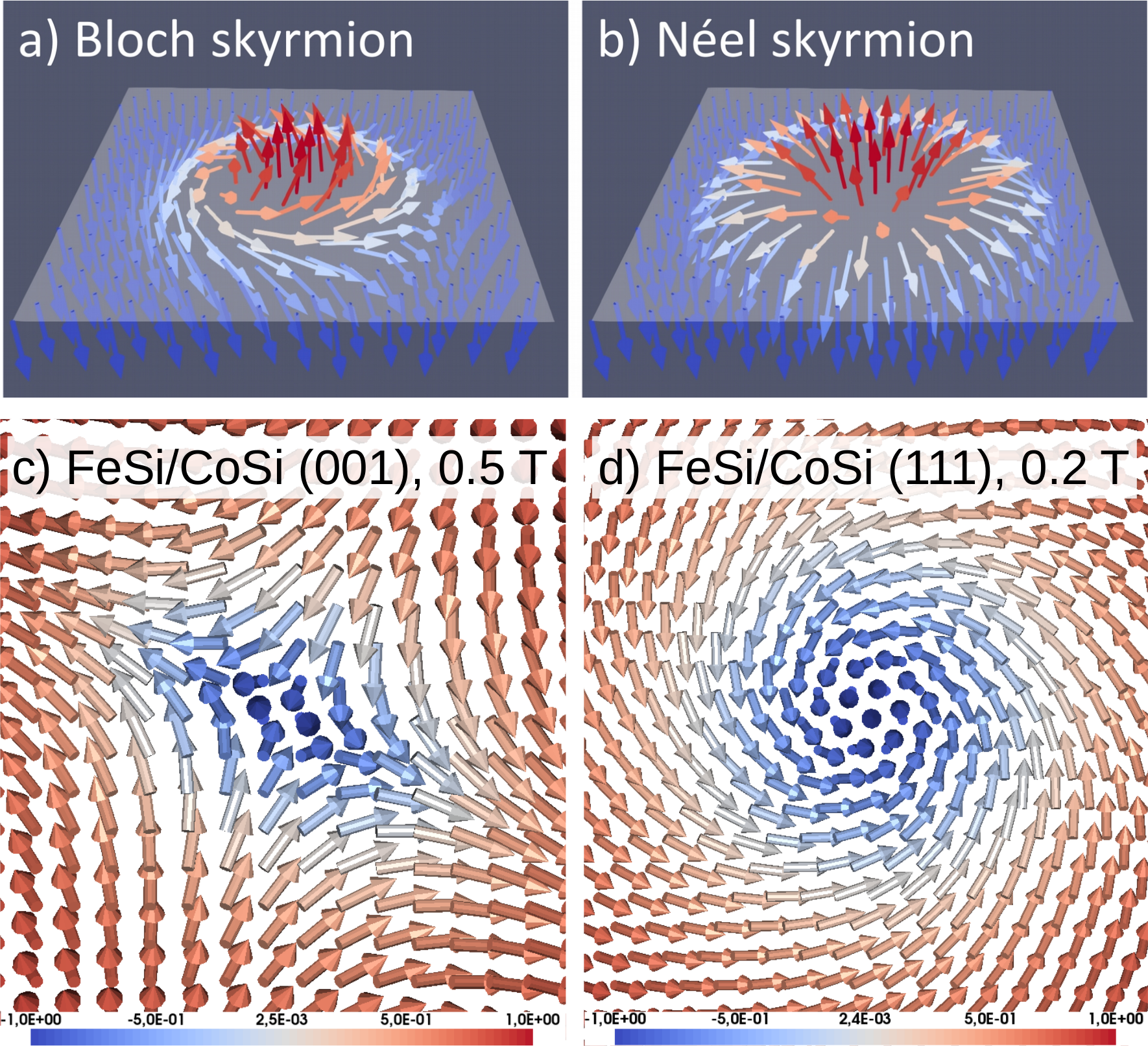}
}
\vspace{-10pt}
\caption{Schematic spin configurations of a) Bloch and b) N\'eel skyrmions usually found in topological magnets. Examples of textures that we predict for B20 FeSi/CoSi multilayers based on atomistic spin dynamics in applied magnetic field: c)~antiskyrmion ($d=\unit[7]{nm}$) and d) intermediate skyrmion ($d=\unit[12]{nm}$). The color code shows the $z$-component of the magnetization.}
\label{f:skyrmions}
\end{figure}

A large part of information is nowadays stored magnetically in hard disk drives where oppositely polarized ferromagnetic domains encode the ``1'' and ``0'' bits. While this simple idea has significantly shaped computer technology during the last decades, it has almost reached its limit in terms of information density and data access time [\onlinecite{Wood2000}]. New ways of magnetic information storage was suggested in 2013~[\onlinecite{Fert2013}] and later on extended in Ref.[\onlinecite{Jena2020}] where the idea is to use nanoscale topological magnetic objects (skyrmions, Fig.~\ref{f:skyrmions}a,b) formed by winding atomic spins in a racetrack to store and process information. Topological magnets can be used not just for storage but also for new types of computing (neuromorphic [\onlinecite{Huang2017},\onlinecite{Bourianoff2018}] and stochastic [\onlinecite{Pinna2018}]). Understandably, searching for new systems of that kind is an on-going active effort in the research community (for a review, see [\onlinecite{Goebel2021}]).

The B20 compound MnSi was the first observed bulk, solid state system with magnetic skyrmions [\onlinecite{Muehlbauer2009}]. Due to the cubic symmetry of its crystal structure, the Dzyaloshinskii-Moriya interaction (DMI) is of well-known bulk type which corresponds to the micromagnetic energy density
\begin{equation}
    \varepsilon_\mathrm{bulk} = D\,\vec{m}\cdot(\vec{\nabla}\times\vec{m}),
    \label{e:bulk_DMI}
\end{equation}
where $\vec{m}=\vec{m}(\vec{r})$ is a continuous function describing the magnetization in different points of space and $D$ is the strength of the DMI. This kind of DMI stabilizes Bloch-type skyrmions where spins rotate perpendicular to the radial direction (Fig.~\ref{f:skyrmions}a). This kind of skyrmion is observed in most bulk topological magnets, except for lacunar spinels [\onlinecite{Kezsmarki2015},\onlinecite{Kanazawa2017}].

Skyrmions have also been found in transition metal multilayers, e.g.~Ir/Fe/Co/Pt [\onlinecite{Soumyanarayanan2017}], Fe/Ir(111) [\onlinecite{Heinze2011}] and Pd/Fe/Ir(111) [\onlinecite{Romming2013}], where the crystal symmetry is quite different. In many cases, these multilayers have a $C_{3\nu}$ symmetry which leads to the so-called interfacial DMI (for details see Ref.~\onlinecite{Bogdanov2001}, Eqn.~(8)) described by the micromagnetic interfacial energy density
\begin{equation}
    \varepsilon_\mathrm{int} = D \left[ m_x \frac{\partial m_z}{\partial x} - m_z \frac{\partial m_x}{\partial x} + m_y \frac{\partial m_z}{\partial y} - m_z \frac{\partial m_y}{\partial y} \right].
    \label{e:interfacial_DMI}
\end{equation}
This interaction, when sufficiently strong, stabilizes N\'eel skyrmions where spins rotate along the radial direction (Fig.~\ref{f:skyrmions}b). One can notice that the compounds building these nanoscale multilayers have no DMI in the bulk, and non-zero DMI is induced by atomically-sharp interfaces. Other interface symmetry ($C_4$) compatible with skyrmions have been analyzed as well for oxides [\onlinecite{Li2014}] and Mn/W(001)-based multilayers [\onlinecite{Nandy2016}].

The materials science community with a focus on systems with non-trivial magnetic topology has the ambition to find suitable materials with optimal size and stability of skyrmions to allow for technological applications. Theoretical calculations based on density functional theory offer a fast and rather reliable route to achieve this goal, although most successful predictions have been for magnetic materials where topology is not considered important. Notable examples here is the tunneling magnetoresistance (TMR) effect that was predicted [\onlinecite{Butler2001}] before it was confirmed experimentally [\onlinecite{Bowen2001}]. Other examples can be found in thin films and surfaces, where predictions of the enhancement of both spin- and orbital magnetic moments [\onlinecite{Jepsen1980}, \onlinecite{Eriksson1991}] preceded experiments [\onlinecite{Lau2002}, \onlinecite{Tischer1995}]. Interestingly, new magnetic materials may be formed by a combination of elements that are nonmagnetic; ZrZn$_2$ and UGe$_2$ are perhaps the most well known example of this and, furthermore, they also exhibit ferromagnetism in combination with superconductivity [\onlinecite{Pfleiderer2001}]. 

The analysis above leads to the main idea of this paper, which is to consider skyrmion-magnetism in materials that are not even magnetic as bulk, but where the consideration of multilayers may induce magnetism and even topological magnetism. Examples of materials relevant here are pure FeSi [\onlinecite{Aeppli1992}] and CoSi [\onlinecite{Wernick1972}]. Neither compound shows long-range magnetic order. However, magnetism is found in the alloy Fe$_{1-x}$Co$_{x}$Si, something which is due to a Stoner instability induced by the change of the electron count combined with peaks in the density of states. 

The first example that we consider is the FeSi/CoSi multilayer. Based on the crystal symmetry considerations, its constituents have a potential to host DMI in the bulk phase, at least when they are alloyed with each other to form the magnetic bulk Fe$_{1-x}$Co$_{x}$Si compound, which is known to host Bloch skyrmions [\onlinecite{Yu2010}] and has a bulk-type DMI. This is due to the cubic B20 crystal symmetry, which is similar to the MnSi compound. For comparison, we also investigate another system, the FeSi/FeGe multilayer, where FeGe already has significant magnetic moments and non-zero DMI in the bulk. The underlying hypothesis of our work is that the interfaces in B20 multilayers can induce magnetism in the FeSi/CoSi system and overall a more complex DMI compared to the bulk B20 systems, and that this DMI will be tunable in a wide range in terms of magnitude and character, for example, by the thickness of the multilayer and interface structure.

It is necessary to note that thin films of FeGe were considered in the literature [\onlinecite{Yu2011}] and a huge enhancement of skyrmionic stability has been found when the film thickness is below the skyrmion lattice spacing $\sim\unit[70]{nm}$. However, for the thicknesses $\unit[(15-75)]{nm}$ that they considered in Ref. [\onlinecite{Yu2011}], the atomistic interface effects do not yet play any role. In the theoretical work presented here, we focus on the FeSi/CoSi and FeSi/FeGe multilayers with a thickness of each component of just a few nanometers, which are not studied experimentally yet. Our goal is to predict significant changes in magnetic properties, that potentially could lead to experimental realization of new types of topological magnetic textures.

To address the complex magnetic behavior of B20 multilayers that we propose in this work, we use a multiscale approach which describes the system step-by-step on increasing length scales (detailed account can be found in our recent works [\onlinecite{Borisov2022},\onlinecite{Borisov2023}]). The first step is the calculation of the electronic properties using density functional theory~[\onlinecite{Hohenberg1964}], available in the full-potential Linear Muffin-Tin Orbital RSPt software [\onlinecite{Wills1987},\onlinecite{Wills2010}]. Generalized-gradient approximation in the PBE parameterization [\onlinecite{PBE1996}] is used, similarly to our previous work on bulk doped B20 compounds [\onlinecite{Borisov2022}]. The FeSi/CoSi multilayer is modelled by a supercell with $n$ unit cells of CoSi in the $z$-direction and $m$ unit cells of FeSi attached to CoSi. Periodic boundary conditions along $x$, $y$ and $z$-directions are imposed, meaning that the chosen model represents a superlattice with repeated (FeSi)$_m$/(CoSi)$_n$ block. For briefness, we will also refer to this kind of structures as $m/n$-superlattices.

We consider two different interface orientations, [001] and [111]. The former is constructed directly by stacking the bulk unit cells of both components along one of the cubic lattice vectors, while the [111]-structure is obtained using the functionality of the Atomic Simulation Environment [\onlinecite{ASE2002},\onlinecite{ASE2017}] which allows to define surfaces with different Miller indices. 
In all cases, the crystal structure is fully optimized using the VASP code [\onlinecite{Kresse1996}]. The same procedure is applied in case of the [001]-FeSi/FeGe multilayers. 
It is worth noting that the literature values of the lattice parameters of the corresponding bulk compounds are $\unit[4.485]{\AA}$ (FeSi), $\unit[4.45]{\AA}$ (CoSi), and $\unit[4.70]{\AA}$ (FeGe). This means that the FeSi/CoSi multilayers are relatively weakly strained due to a small lattice mismatch of 0.8\% between FeSi and CoSi. The FeSi/FeGe superlattices show a larger lattice mismatch of 4.8\%.

To evaluate the changes of magnetic interactions between the Fe and Co moments near the interfaces compared to the bulk case, we calculated the Heisenberg and DM magnetic interactions using the Liechtenstein-Katsnelson-Antropov-Gubanov (LKAG) approach [\onlinecite{LKAG1987}] (review in \onlinecite{Szilva2023}), available in the RSPt software [\onlinecite{Wills1987},\onlinecite{Wills2010}]. The system is mapped then on an effective spin model:
\begin{equation}
    H = -\sum\limits_{j\neq i} J_{ij} (\vec{S}_i \cdot \vec{S}_j) -\sum\limits_{j\neq i} \vec{D}_{ij} \cdot (\vec{S}_i \times \vec{S}_j).
    \label{e:spin_model}
\end{equation}
The advantage of the LKAG approach is that interactions up to a distance of several lattice parameters, which contains several hundred interaction parameters, can be calculated without using large supercells, as for example in the total energy fitting method \textbf{[ref]}.

With this information, we simulate the magnetic properties of the B20 multilayers at finite temperature and applied magnetic field using atomistic spin dynamics (ASD) and Monte Carlo simulations, available in the \textsc{UppASD} code~[\onlinecite{uppasd},\onlinecite{Eriksson2017}]. The ASD simulations are based on the Landau-Lifshitz-Gilbert equation [\onlinecite{Landau1935},\onlinecite{Gilbert2004}]:
\begin{equation}
    \frac{\partial \vec{m}_i}{\partial t} = -\frac{\gamma}{1 + \alpha^2} \left[ \vec{m}_i \times \vec{B}_i + \frac{\alpha}{m}\,\vec{m}_i \times (\vec{m}_i \times \vec{B}_i) \right],
    \label{e:LLG_equation}
\end{equation}
which describes the time-evolution of the magnetization $\vec{m}_i$ of a given spin $i$ under the influence of the effective field $\vec{B}_i$. The later is determined from the spin model (Eqn.\ref{e:spin_model}) where, to a first approximation, we assume a zero on-site anisotropy. As is usual for Langevin dynamics, a random field proportional to $\sqrt{\alpha\, T}$ was added to $\vec{B}_i$ to simulate finite-temperature fluctuations, and the damping constant $\alpha$ is set to $1.0$ to enable a fast convergence to the (quasi)-equilibrium state. We perform these ASD simulations for a $(500\times 500\times 1)$ supercell with periodic boundary conditions, which contains $10^6$ spins for the (FeSi)$_3$/(CoSi)$_3$ and $2\cdot 10^6$ spins for the (FeSi)$_4$/(CoSi)$_1$ superlattice. The slab with less amount of CoSi contains more spins, because the whole CoSi becomes magnetic, while in thicker CoSi layer only the interfaces are magnetic. The supercell cross-sections correspond to a simulated region of around $\unit[(220\times 220)]{nm^2}$ which allows to accommodate spin spiral and skyrmion phases.

\begin{figure}[h]
{\centering
\includegraphics[width=0.95\columnwidth]{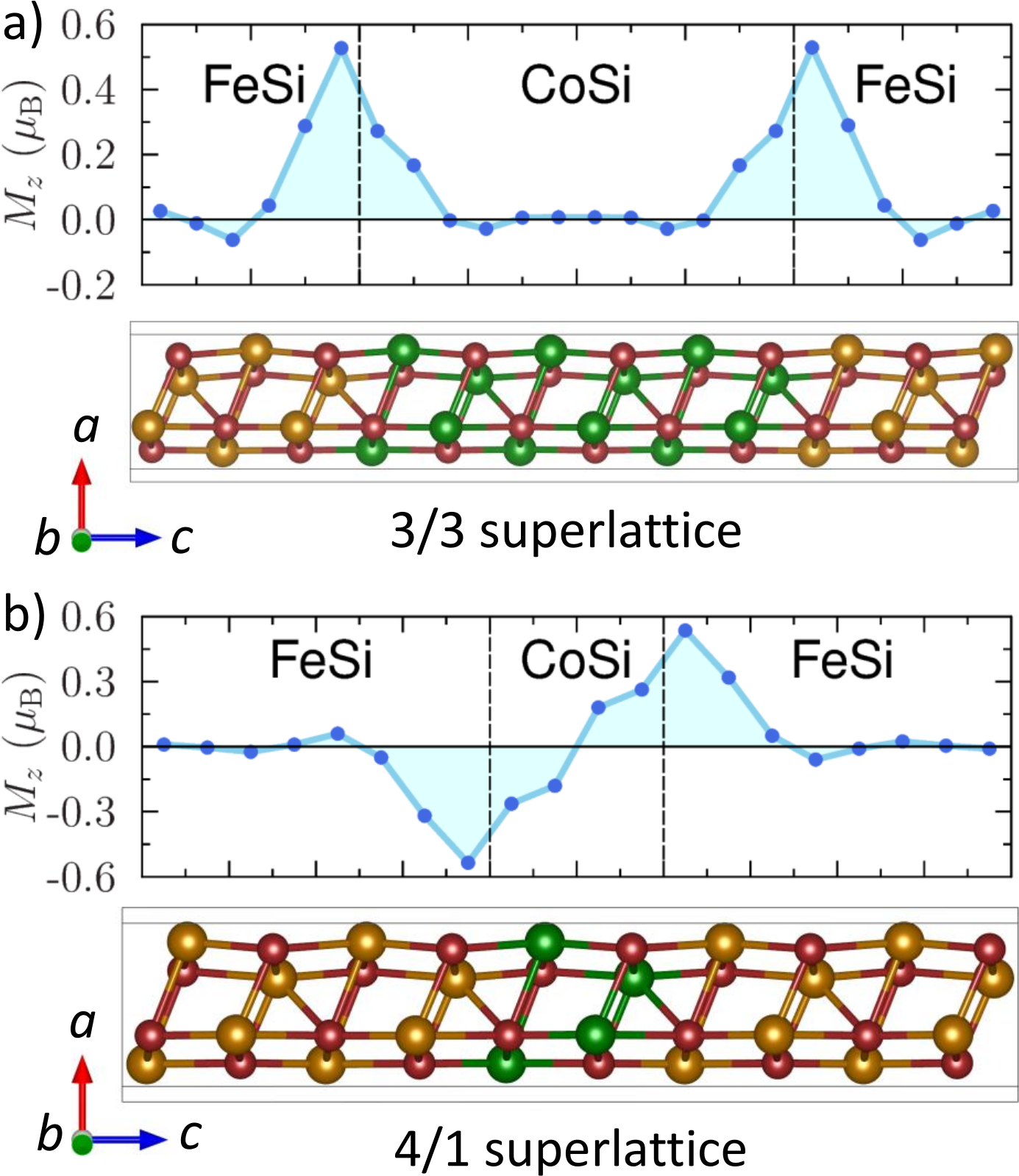}
}
\caption{Layer-resolved magnetic moments on the Fe and Co sites in the a) (FeSi)$_3$/(CoSi)$_3$ and b) (FeSi)$_4$/(CoSi)$_1$ superlattices; the interfaces are marked by vertical dashed lines. Below each plot, a structural sketch of the system is shown, where Fe (light-brown) and Co (green) sites are connected by Si (red-brown).}
\label{f:FeSi_CoSi_superlattices}
\end{figure}

\vspace{5pt}
\textit{$\mathrm{(FeSi)}_m/\mathrm{(CoSi)}_n$ superlattices.} For these systems, we find the formation of finite magnetic moments on Fe and Co within a few atomic layers near the interfaces due to a changed electron count, similarly to the bulk Fe$_{1-x}$Co$_{x}$Si alloy. Further away from the interfaces, both FeSi and CoSi show no intrinsic magnetic moments, which is consistent with the magnetism for the bulk counterparts. This is illustrated in Fig.~\ref{f:FeSi_CoSi_superlattices} on the example of the 3/3-superlattice where the induced moments are below $\unit[0.6]{\mu_\mathrm{B}}$ and localized around each FeSi/CoSi interface. Even for the 2/2-superlattice with smaller thickness of FeSi and CoSi we find that the middle parts of the slabs are basically non-magnetic and the overall profile of magnetization near the interfaces is very similar to that of the 3/3-superlattice (data not shown). However, for the 4/1-superlattice, where the CoSi layer is just 1 unit cell thick, we find that the whole CoSi becomes magnetic but the two interfaces are antiferromagnetically coupled, so the resulting magnetization of the system is zero (Fig.~\ref{f:FeSi_CoSi_superlattices}b). Interestingly, the sums of absolute values of local moments in both superlattices are very similar, in the range $\unit[2.3-2.4]{\mu_\mathrm{B}}$ per surface unit cell, despite different amounts of Fe and Co in both cases. This is because the magnetic properties in the studied structures are dominated by the interfaces, where the local doping due to mixture of Fe and Co leads to a Stoner instability and finite magnetic moments, similarly to the bulk B20 alloy Fe$_{1-x}$Co$_{x}$Si. The AFM ordering in the 4/1-superlattice, however, indicates significantly different magnetic interactions compared to the 3/3-system. As we will see in the following (page 5), the relative strength of the Dzyaloshinskii-Moriya (DM) interaction, which is crucial for topological magnetism, can be also sensitive to the interface structure and be quite different from the bulk counterparts.

\begin{figure}[h]
{\centering
\includegraphics[width=0.99\columnwidth]{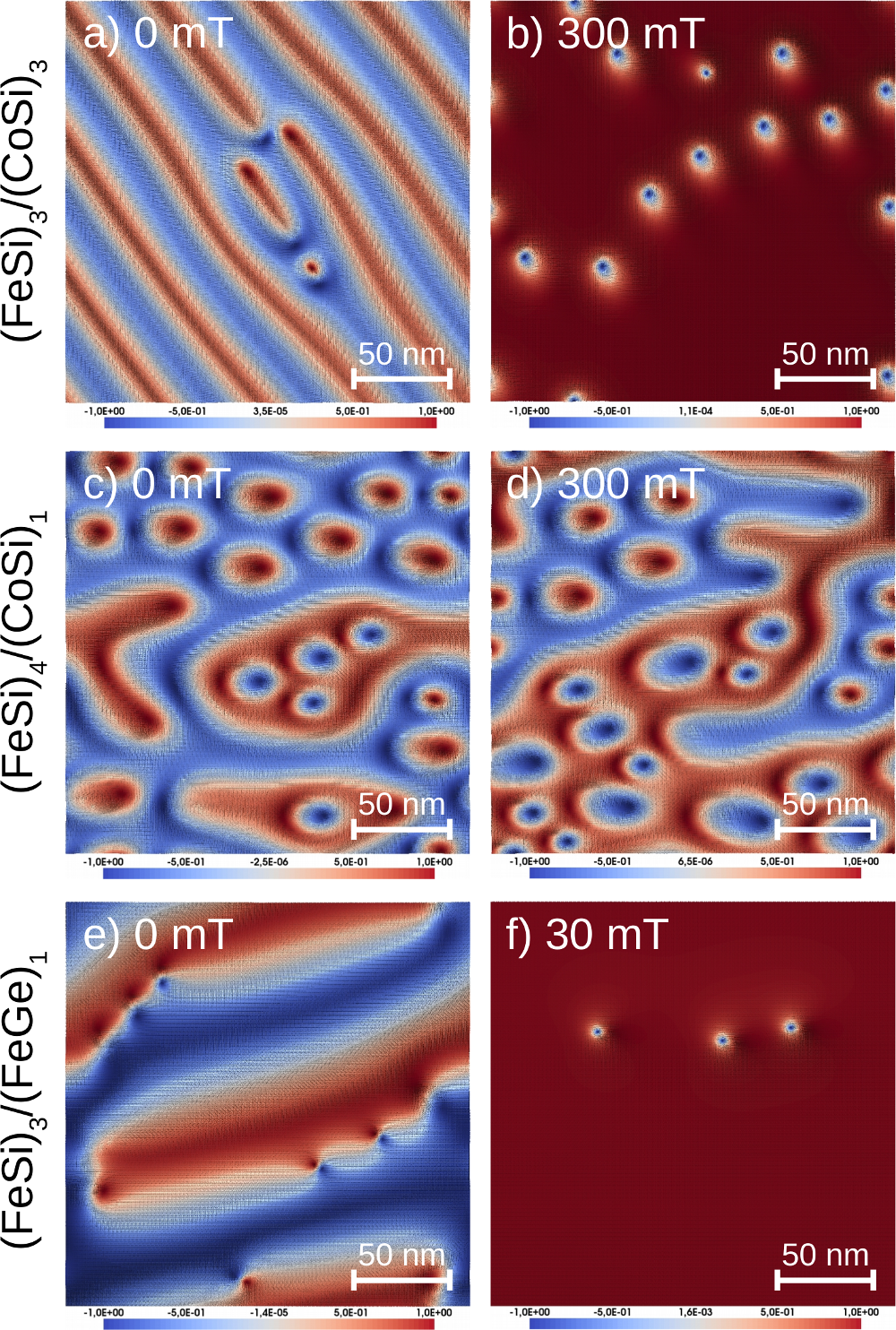}
}
\vspace{-10pt}
\caption{Predicted magnetic ground state of the a,b) (FeSi)$_3$/(CoSi)$_3$, c,d) (FeSi)$_4$/(CoSi)$_1$ and e,f) (FeSi)$_3$/(FeGe)$_1$ superlattices at zero and finite external magnetic field perpendicular to the interfaces from atomistic spin dynamics simulations of annealing from $\unit[300]{K}$ to zero temperature. The size of the simulated region is around $\unit[(220\times 220)]{nm^2}$ and includes $10^6$ spins for a,b) and $2\cdot 10^6$ spins for c-f). Color code depicts the $z$-component of spin, which varies between $-1$ and $+1$.}
\label{f:magnetic_ground_state}
\end{figure}
The simulated annealing procedure, where one goes from high to low temperatures in the ASD simulations, allows to predict the magnetic ground state of the (FeSi)$_3$/(CoSi)$_3$ system at zero external field, which is a spin spiral with a period around $\unit[45]{nm}$, as shown in Fig.~\ref{f:magnetic_ground_state}a. A few topological defects in this spiral phase can be seen in this figure, and their presence may depend on the cooling rate in the annealing simulations. When a finite magnetic field is applied to the system perpendicular to the interfaces ($z$-direction), the spirals start to split (Fig.~\ref{f:spin_textures_SI_1}b,c) and around $\unit[300]{mT}$ one sees the formation of multiple isolated, somewhat distorted (elongated along the [110] direction), antiskyrmions with a size $\sim\!\unit[14]{nm}$ (Fig.~\ref{f:magnetic_ground_state}b) as well as one intermediate skyrmion with a size $\sim\!\unit[8]{nm}$ (Fig.~\ref{f:spin_textures_closeups}a). The 2/2-superlattice shows a very similar behavior (Fig.~\ref{f:spin_textures_SI_1}e-h). Usually, skyrmions have a topological number $\pm 1$, depending on the orientation of the magnetization in the center of the skyrmion, and the helicity is either 0 (N\'eel skyrmion) or $\pi/2$ (Bloch skyrmion). For the intermediate skyrmion, the helicity deviates from these two limits, but is closer to the Bloch limit for the (FeSi)$_3$/(CoSi)$_3$ multilayer, as can be seen from a close-up of this texture (Fig.~\ref{f:spin_textures_closeups}a in the SI). Using the methodology discussed in Ref.[\onlinecite{Kim2020}], we calculate the topological charge of the observed antiskyrmions to be close to $-1$ and of the intermediate skyrmions to be around $+1$. At higher fields around $\unit[400-500]{mT}$ the antiskyrmions become smaller and less distorted ($\sim\!\unit[7-8]{nm}$ in Fig.~\ref{f:skyrmions}c) and, finally, above $\unit[600]{mT}$ the system becomes fully ferromagnetic (FM phase). Monte Carlo simulations for the FM state in zero field allow to estimate the critical temperature $T_\mathrm{C}$ for magnetic ordering to be around $\unit[40]{K}$ (Fig.~\ref{f:M_vs_T}a). It may seem lower than $T_\mathrm{C}$ of bulk Fe$_{0.5}$Co$_{0.5}$Si seen in Fig.~\ref{f:M_vs_T}e, but the later is overestimated by theory compared to the measured values around $\unit[30]{K}$. Based on this, we expect that the thermal stability of [001]-FeSi/CoSi superlattice is similar to the bulk counterpart.

In contrast, for the (FeSi)$_4$/(CoSi)$_1$ superlattice we find antiferromagnetically (AFM) coupled interfaces with zero net magnetization, as mentioned already on page 2 and shown in Fig.~\ref{f:FeSi_CoSi_superlattices}b. However, the distribution of magnetization near each interface contains interesting features, shown in Fig.~\ref{f:magnetic_ground_state}c,d for one of the interfaces (the other interface nearby has opposite magnetization but the same distribution in real space). At zero field, we find a number of distorted antiskyrmions $\sim\!\unit[18-37]{nm}$ with topological charge $-1$ embedded in locally ferromagnetic (FM) domains and their number changes in an applied magnetic field, but their average size is barely affected and the total magnetization of each interface is around or below $\unit[0.15]{\mu_\mathrm{B}}$ per $(500\times500\times1)$ supercell, meaning that it is magnetically almost compensated. In a way, this FeSi/CoSi system represents a synthetic antiferromagnet with topological magnetic textures. Due to the strong AFM coupling between the interfaces, the system is not easy to polarize by external magnetic field nor is it easy to manipulate the antiskyrmions, compared to the 3/3- and 2/2-superlattices. In the simulations of the AFM coupled system, it is very difficult to induce any significant ferromagnetic polarization, even when using fields up to $\unit[1]{T}$. This is due to the strong AFM exchange between the two neighboring FeSi/CoSi interfaces, that dominates any external field in the simulations. This may offer unique advantages in terms of stability of the topological magnetic states that based on the calculations put forth here, would show unusual resilience to perturbations from a magnetic field. 
Based on the temperature-dependent simulations of the collinear AFM phase (Fig.~\ref{f:M_vs_T})c, we expect that antiskyrmions in the 4/1-superlattice can be stable below $\unit[90]{K}$.

\begin{figure}[h]
{\centering
\includegraphics[width=0.99\columnwidth]{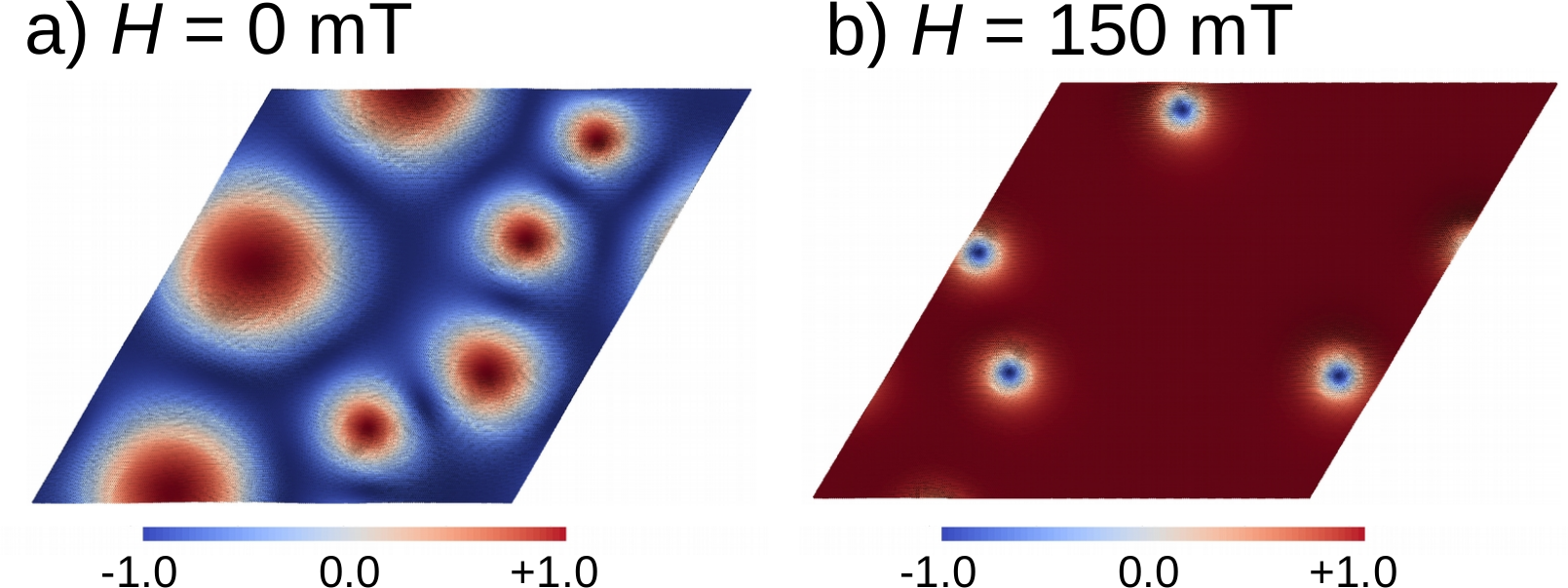}
}
\vspace{-10pt}
\caption{Predicted magnetic ground state of the [111]-oriented (FeSi)$_3$/(CoSi)$_3$ superlattice at a) zero and b) finite external magnetic field $H=\unit[150]{mT}$ along the [111]-direction from atomistic spin dynamics simulations of annealing from $\unit[300]{K}$ to zero temperature. The simulated region has a side length of around $\unit[157]{nm}$ and $60^\mathrm{o}$ angle in-between and includes $8.75\cdot10^5$ spins. In both plots, the point-like topological objects are intermediate skyrmions. Color code depicts the $z$-component of spin, which varies between $-1$ and $+1$.}
\label{f:magnetic_ground_state_111}
\end{figure}

We also considered the [111]-oriented (FeSi)$_3$/(CoSi)$_3$ multilayer where, similarly to the [001] interfaces, magnetism emerges at the interfaces, while the interfaces are separated by almost non-magnetic regions. From the ASD simulations using a $(250\times250\times1)$ supercell with 875,000 atomic moments we find the zero-field ground state containing intermediate skyrmions with topological charge $+1$ (Fig.~\ref{f:magnetic_ground_state_111}a), which is in contrast to the zero-field spin-spiral phase of the [001] interface. In external magnetic field between $\unit[(100-200)]{mT}$, these skyrmions become more compact (Fig.~\ref{f:magnetic_ground_state_111}b); their diameter varies monotonically between $\unit[24]{nm}$ at $H=\unit[100]{mT}$ and $\unit[12]{nm}$ at $H=\unit[200]{mT}$. At fields above $\unit[300]{mT}$, the system becomes fully ferromagnetic, and the critical temperature of the FM phase in zero field from Monte Carlo simulations is similar to the [001]-oriented superlattice (Fig.~\ref{f:M_vs_T}a).

Next, we describe the results for the $\mathrm{(FeSi)}_m/\mathrm{(FeGe)}_n$ superlattices. The main difference between the FeSi/FeGe and FeSi/CoSi systems is that FeGe already shows finite magnetic moments in the bulk, while CoSi does not. Near the [001]-oriented interface, however, we find that the FeGe magnetization is reduced considerably compared to the bulk material (Fig.~\ref{f:FeSi_FeGe_superlattices} in the SI), which reduces considerably the critical temperature for the 3/1-superlattice compared to bulk, while the 3/3-superlattice is more reminiscent of bulk (compare Figs.~\ref{f:M_vs_T}b,d,f). Also, non-zero moments $\sim\!\unit[0.3]{\mu_B}$ are induced in the neighboring FeSi layer; these are, however, not included in our spin model due to their much smaller magnitude compared to the Fe moments in FeGe. It should be emphasized that these changes are just within one atomic layer near the interface and are observed both in 3/3- and 3/1-superlattices. For the [111]-oriented (FeSi)$_3$/(FeGe)$_1$ multilayer, we could not stabilize any sizeable moments on Fe sites in the RSPt calculations.

The ASD annealing simulations for the (FeSi)$_3$/(FeGe)$_1$ superlattice using a $(500\times 500\times 1)$ supercell containing  $10^6$ spins predict a spin spiral phase at zero field with a comparably large spatial period of around $\unit[100]{nm}$ (Fig.~\ref{f:magnetic_ground_state}e). Interestingly, this system appears to be more sensitive to external magnetic field (Fig.~\ref{f:magnetic_ground_state}f) than the other systems that we studied, since already for $H=\unit[40]{mT}$ it becomes fully ferromagnetic. At a slightly smaller field $H=\unit[30]{mT}$ we see isolated Bloch skyrmions with topological charge $+1$ (Fig.~\ref{f:magnetic_ground_state}f and Fig.~\ref{f:spin_textures_closeups} in the SI), which are also known for bulk FeGe in a similar range of fields (schematic view in Fig.~\ref{f:skyrmions}a). In our multilayer, however, the skyrmions are much more compact ($\sim\!\unit[8]{nm}$) than in bulk FeGe where the characteristic magnetic length scale is $\unit[70]{nm}$ [\onlinecite{Lebech1989,Uchida2008,Yu2011}]. Interestingly, despite marked changes of the DM matrix compared to bulk FeGe (see Table~I in the SI), the skyrmions in this (FeSi)$_3$/(FeGe)$_1$ (001) multilayer are still of Bloch character.

To analyze the origin of the magnetic topology observed in our simulations, we calculate and compare the micromagnetic parameters of the studied multilayers, which is easier than analyzing the corresponding atomistic interactions for hundreds of atomic neighbors. We do not perform actual micromagnetic simulations in this work, because the atomistic spin dynamics is a more accurate and straightforward way to simulate these systems with several magnetic atoms per unit cell and quite small skyrmions of the order of several nanometers. Nevertheless, we make a note on how to calculate the micromagnetic parameters and the effective field for micromagnetic simulations in Sec.~\ref{t:calc_micro} and \ref{t:micro_field} of the SI, which can be useful as a general reference.

The micromagnetic quantities $A$ and $D_{\alpha\beta}$ for the different B20 superlattices are collected in Table~I in the SI and give an overall impression of the strength of Heisenberg and Dzyaloshinskii-Moriya interactions in different multilayers that we studied. Here, the energy scale of the DMI is characterized by the sum $\left\langle D \right\rangle = \sum\limits_{\alpha\beta} D_{\alpha\beta}$ of different components of the micromagnetic DM matrix. First, we notice that the 4/1-superlattice appears to show a somewhat smaller $A/\left\langle D \right\rangle$ ratio (i.e.~more pronounced DM interaction) compared to the 3/3-superlattice. We argue that this is related to the smaller distance between the interfaces in the 4/1-system, which probably enhances their symmetry breaking effect and increases the DM interaction. Importantly, for bulk Fe$_{0.5}$Co$_{0.5}$Si (see Table~I in SI) we find weaker bulk-type DMI and $A/\left\langle D \right\rangle$ ratio which is almost an order of magnitude larger than the corresponding values for the FeSi/CoSi superlattices, suggesting that nanoscale B20 multilayers can be, indeed, more promising in terms of topological magnetism than the bulk counterparts. For FeSi/FeGe, the increase of DMI compared to bulk is less pronounced, but the structure of the DM matrix is changed and we find rather small Bloch skyrmions in a narrow field range. The important finding of our work is that the B20 multilayers show a much more complicated character of the DM matrix compared to the diagonal DM matrix of bulk B20 compounds (see Table~I in the SI) or simple $D_{xy}$-type DM matrix of transition metal multilayers, leading to stabilization of spin textures beyond N\'eel or Bloch skyrmions (Fig.~\ref{f:skyrmions}).

In conclusion, the B20 multilayers, studied here using state-of-the-art first principles methods, show a potential for interesting topological magnetism, that can be different both from that of the bulk B20 compounds and many other metallic multilayers. Both the interface orientation and layer thicknesses appear to change the character of magnetism and topological textures, which include antiskyrmions and intermediate skyrmions coexisting near FeSi/CoSi interfaces and Bloch skyrmions in FeSi/FeGe multilayer, all in the range of sizes $\sim\!\unit[(7-37)]{nm}$. For most systems, we find that to observe these textures it is not necessary to have atomically thin layers of B20 compounds; instead, a single interface between two relatively thick layers (e.g.~FeSi and CoSi) can be used for that purpose. For technological applications we highlight the (FeSi)$_4$/(CoSi)$_1$  system, since it seems to be particularly stable with respect to external magnetic fields and the AFM skyrmions that we observe there should have zero skyrmion Hall effect, based on previous work \cite{Zhang2016}. Our theoretical predictions call for experimental verifications and motivate studies of similar systems where non-zero magnetic moments and significant chiral magnetic interactions are induced by nanoscale interfaces.

\textit{Acknowledgements.} This work was financially supported by the Knut and Alice Wallenberg Foundation through grant numbers 2018.0060, 2021.0246, and 2022.0108, and G\"oran Gustafsson Foundation (recipient of the ``small prize'': Vladislav Borisov). Olle Eriksson and Anna Delin acknowledge support from the Wallenberg Initiative Materials Science for Sustainability (WISE) funded by the Knut and Alice Wallenberg Foundation (KAW). 
AD also acknowledges financial support from the Swedish Research Council (Vetenskapsrådet, VR), Grant No. 2016-05980 and Grant No. 2019-05304.
Olle Eriksson also acknowledges support by the Swedish Research Council (VR), the Foundation for Strategic Research (SSF), the Swedish Energy Agency (Energimyndigheten), the European Research Council (854843-FASTCORR), eSSENCE and STandUP. The computations/data handling were enabled by resources provided by the Swedish National Infrastructure for Computing (SNIC) at the National Supercomputing Centre (NSC, Tetralith cluster) partially funded by the Swedish Research Council through grant agreement no.\,2018-05973 and by the National Academic Infrastructure for Supercomputing in Sweden (NAISS) at the National Supercomputing Centre (NSC, Tetralith cluster) partially funded by the Swedish Research Council through grant agreement no.\,2022-06725. Structural sketches in Fig.~\ref{f:FeSi_CoSi_superlattices} have been produced by the \textsc{VESTA3} software \cite{vesta}. Fig.~\ref{f:skyrmions}a,b was produced by the \textsc{Paraview} software \cite{paraview}.

\newpage
\onecolumngrid

\renewcommand{\figurename}{Fig.}
\renewcommand{\thefigure}{S\arabic{figure}}
\setcounter{figure}{0}

\begin{center}
    {\large \textbf{Supplemental Information for the manuscript}\\[7pt]
    \textbf{``Tunable topological magnetism in superlattices of nonmagnetic B20 systems''}
    }

    \vspace{7pt}
    {\large by Vladislav Borisov,$^1$ Anna Delin,$^{2,3,4}$ Olle Eriksson$^{1,5}$}

    \vspace{7pt}
    $^1$Department of Physics and Astronomy, Uppsala University, Box 516, SE-75120 Uppsala, Sweden

    $^2$Department of Applied Physics, School of Engineering Sciences, KTH Royal Institute of Technology, AlbaNova University Center, SE-10691 Stockholm, Sweden

    $^3$Wallenberg Initiative Materials Science for Sustainability (WISE), KTH Royal Institute of Technology, SE-10044 Stockholm, Sweden

    $^4$SeRC (Swedish e-Science Research Center), KTH Royal Institute of Technology, SE-10044 Stockholm, Sweden

    $^5$Wallenberg Initiative Materials Science for Sustainability, Uppsala University, 75121 Uppsala, Sweden
    
    \vspace{7pt}
    (Dated: \today)
\end{center}

\setcounter{section}{0}
\section{Atomistic spin dynamics}

\begin{figure}
\includegraphics[width=0.99\columnwidth]{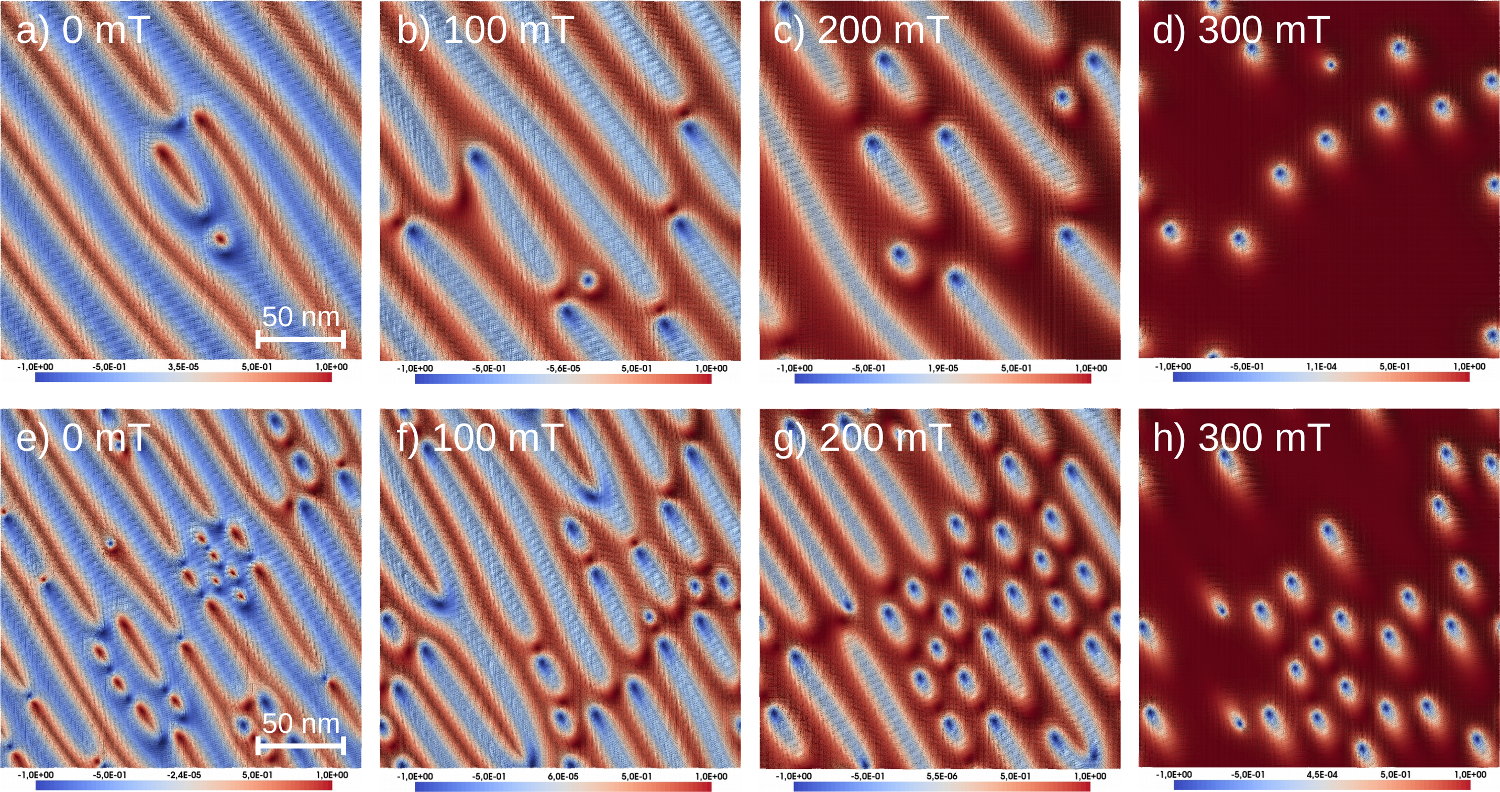}
\caption{Equilibrium spin configurations of the a-d) (FeSi)$_3$/(CoSi)$_3$ and e-h) (FeSi)$_2$/(CoSi)$_2$ superlattices obtained from the atomistic spin dynamics simulations with \textit{ab initio} magnetic interactions. The external magnetic field is applied perpendicular to the interfaces and is varied between $\unit[0]{mT}$ and $\unit[300]{mT}$. The on-site anisotropy is set to zero.}
\label{f:spin_textures_SI_1}
\end{figure}

\begin{figure}
\includegraphics[width=0.99\columnwidth]{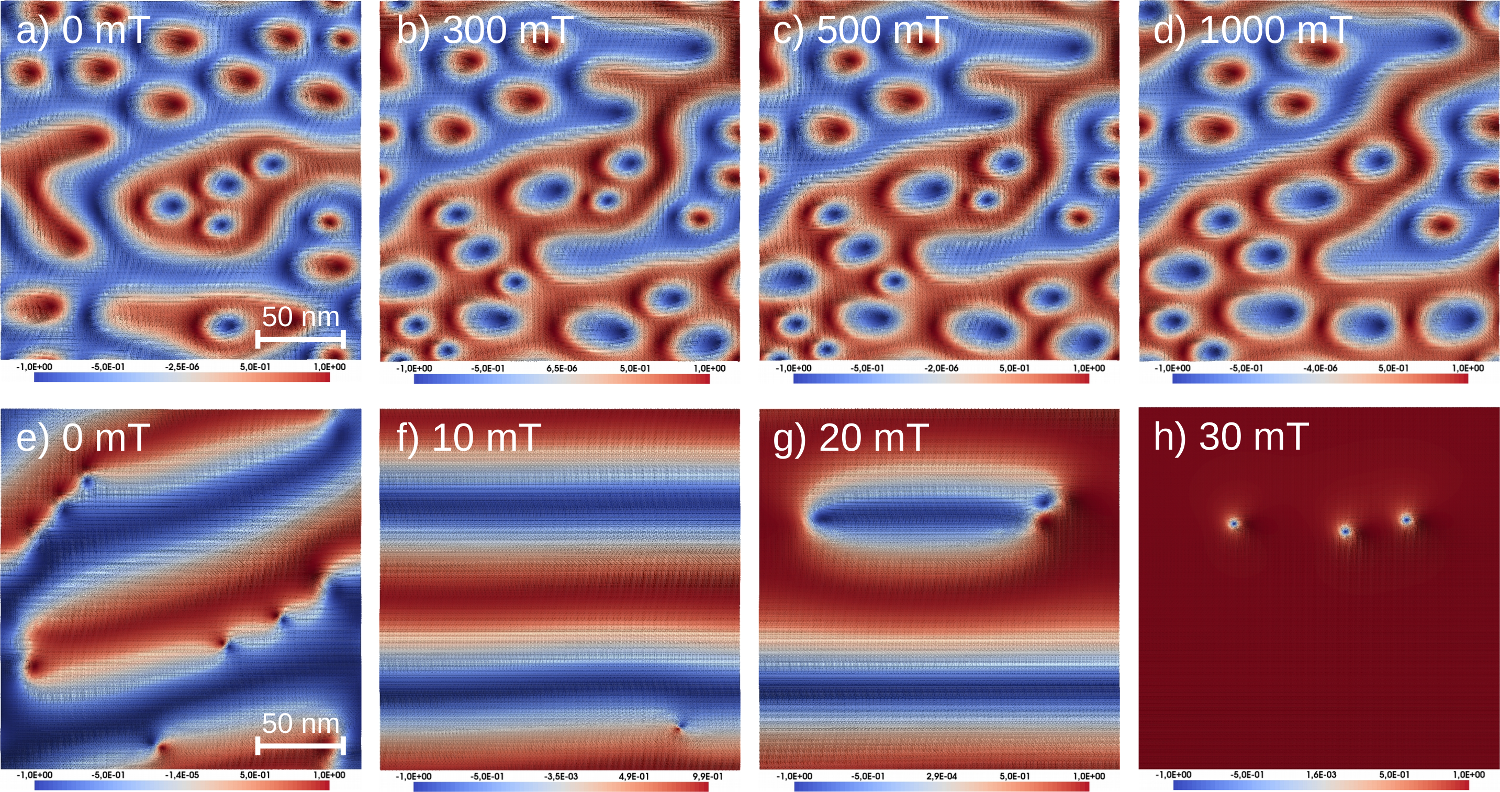}
\caption{Equilibrium spin configurations of the a-d) antiferromagnetic (FeSi)$_4$/(CoSi)$_1$ and e-h) (FeSi)$_3$/(FeGe)$_1$ superlattices obtained from the atomistic spin dynamics simulations with \textit{ab initio} magnetic interactions. The external magnetic field is applied perpendicular to the interfaces and is varied between $\unit[0]{mT}$ and $\unit[1000]{mT}$ for CoSi-based system and between $\unit[0]{mT}$ and $\unit[30]{mT}$ for the FeGe-based system. The on-site anisotropy is set to zero. In case of the (FeSi)$_4$/(CoSi)$_1$ system, the figure shows the spin texture of one of the interfaces; spins at another interface are opposite due to the AFM coupling between the interfaces.}
\label{f:spin_textures_SI_2}
\end{figure}

\begin{figure}
\includegraphics[width=0.99\columnwidth]{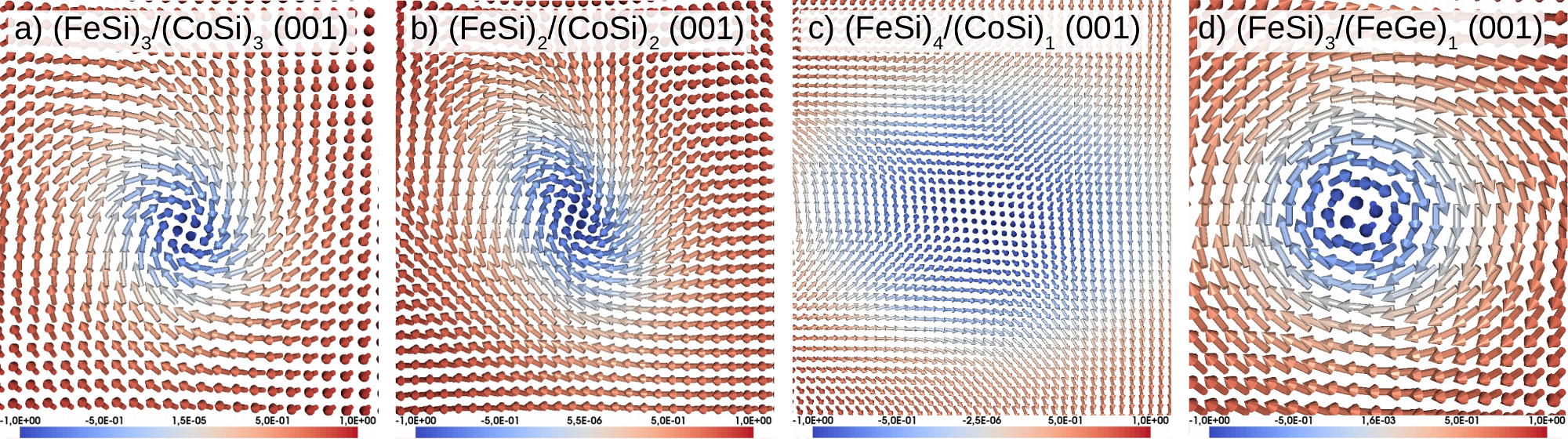}
\caption{Close-ups of the spin configurations of the FeSi/CoSi and FeSi/FeGe multilayers (each atomic spin is represented by an arrow). The field-induced intermediate skyrmions in FeSi/CoSi (a,b) are additional topological objects that coexist with the dominant antiskyrmions (Fig.~\ref{f:skyrmions}c and Fig.~\ref{f:magnetic_ground_state}b). c) Large antiskyrmions $\sim\!\unit[(18-37)]{nm}$ are observed near the AFM coupled FeSi/CoSi interfaces. d) Bloch skyrmions are induced by small magnetic field $\unit[30]{mT}$ in FeSi/FeGe multilayer.}
\label{f:spin_textures_closeups}
\end{figure}

Here, we show more snapshots of ASD simulations for different B20 superlattices. In Fig.~\ref{f:spin_textures_SI_1}, the zero-temperature spin configurations of the 3/3- and 2/2-superlattices of FeSi/CoSi are depicted for different external magnetic fields between $\unit[0]{mT}$ and $\unit[300]{mT}$. In Fig.~\ref{f:spin_textures_SI_2}, we also show the spin configurations of the (FeSi)$_4$/(CoSi)$_1$ and (FeSi)$_3$/(FeGe)$_1$ multilayers for different field ranges. Additional information is shown in Fig.~\ref{f:spin_textures_closeups} for all these systems, with explanation in the figure caption. The snapshots in Figs.~\ref{f:spin_textures_SI_1}--\ref{f:spin_textures_closeups} illustrate almost-equilibrium states reached after several nanoseconds of simulations where the spin texture shows only little changes upon further relaxation on the chosen time scale. Qualitatively, the magnetic textures in 3/3- and 2/2-superlattices of FeSi/CoSi are similar, although the 2/2-system is characterized by a somewhat smaller length scale of spin spirals, consistent with the calculated $A/D$ ratios (Table~I in the SI).

\section{Critical temperatures}

\begin{figure}[h]
{\centering
\includegraphics[width=0.49\columnwidth]{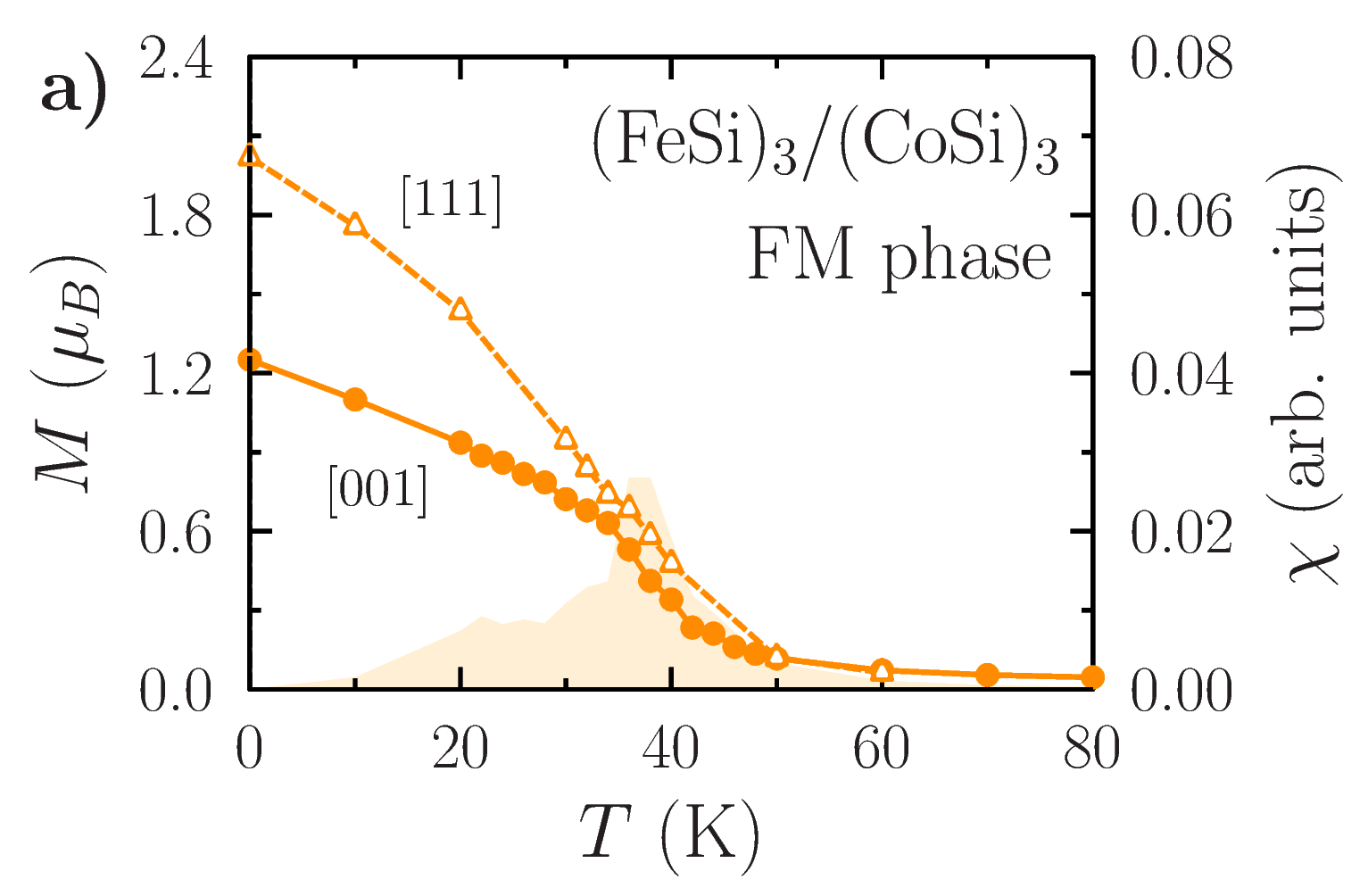}
\includegraphics[width=0.49\columnwidth]{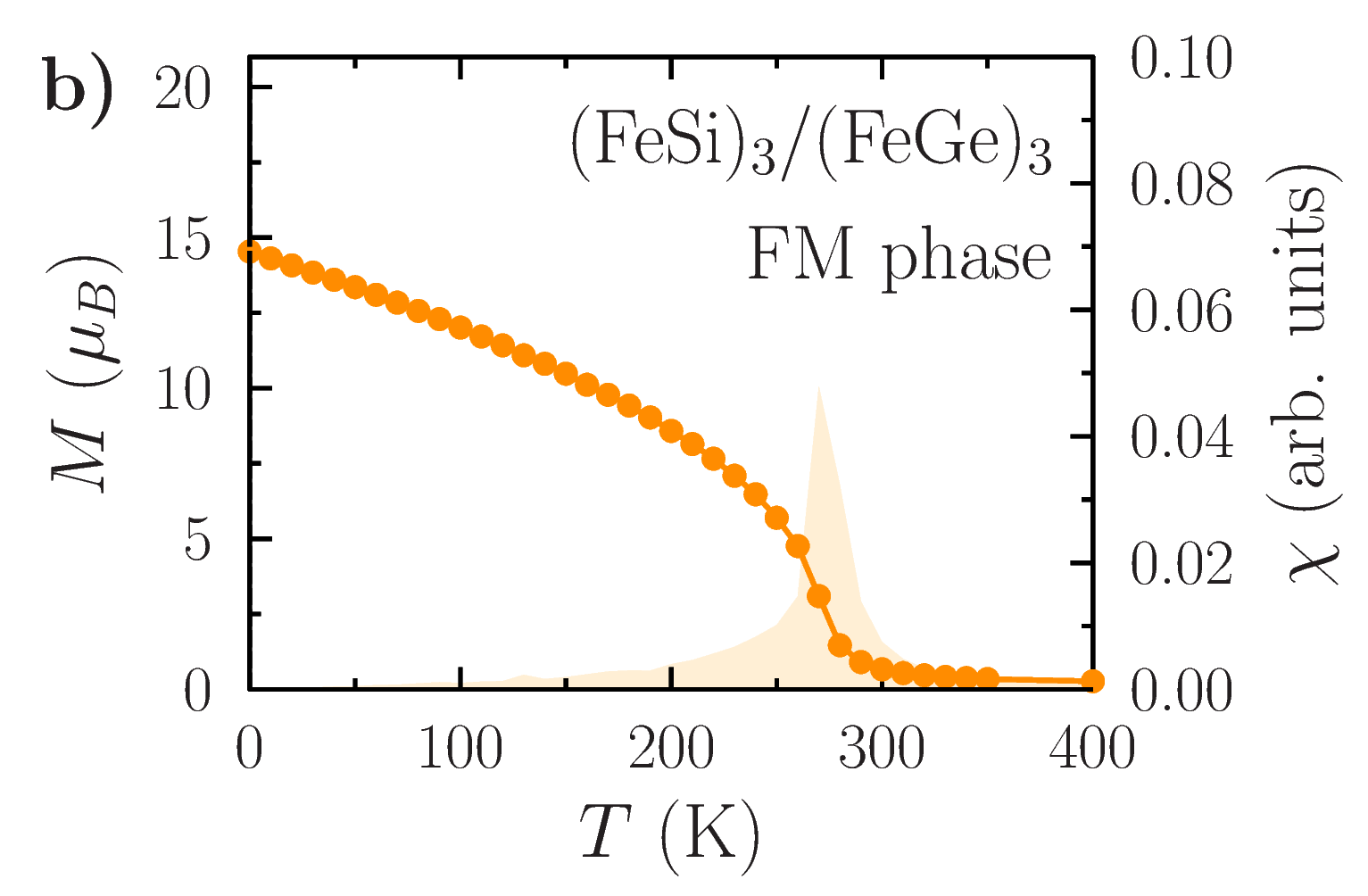}
\includegraphics[width=0.49\columnwidth]{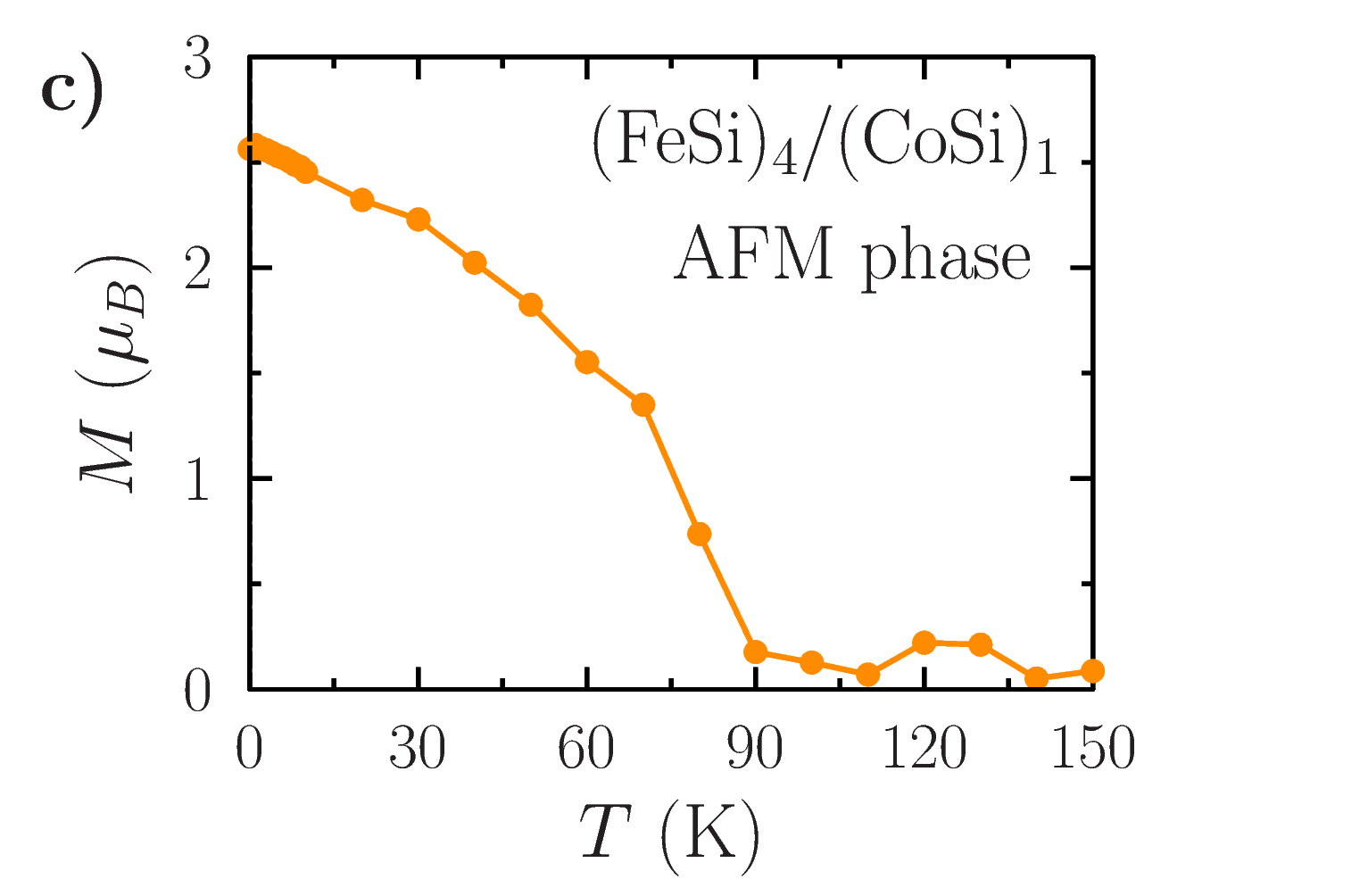}
\includegraphics[width=0.49\columnwidth]{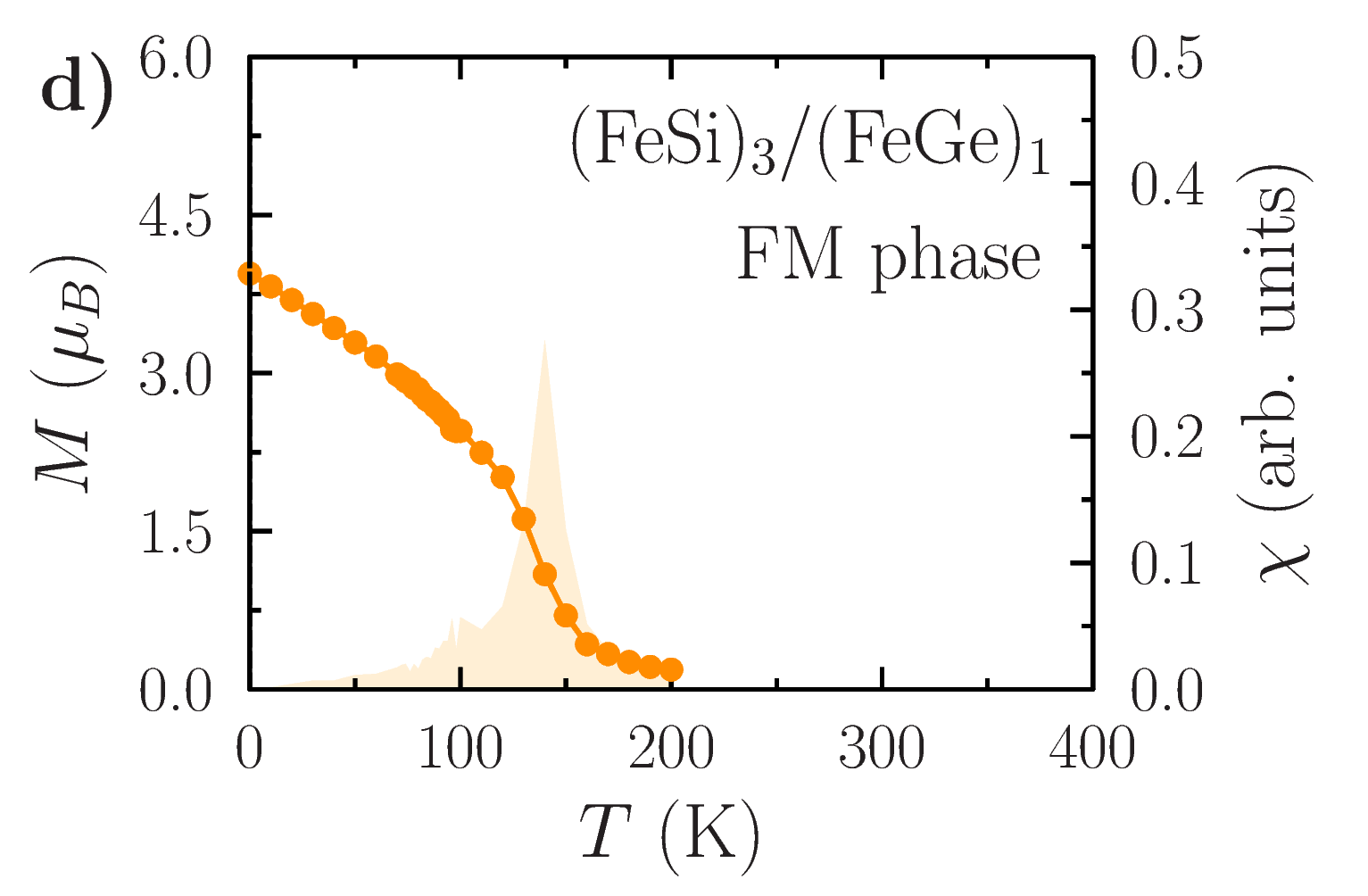}
\includegraphics[width=0.49\columnwidth]{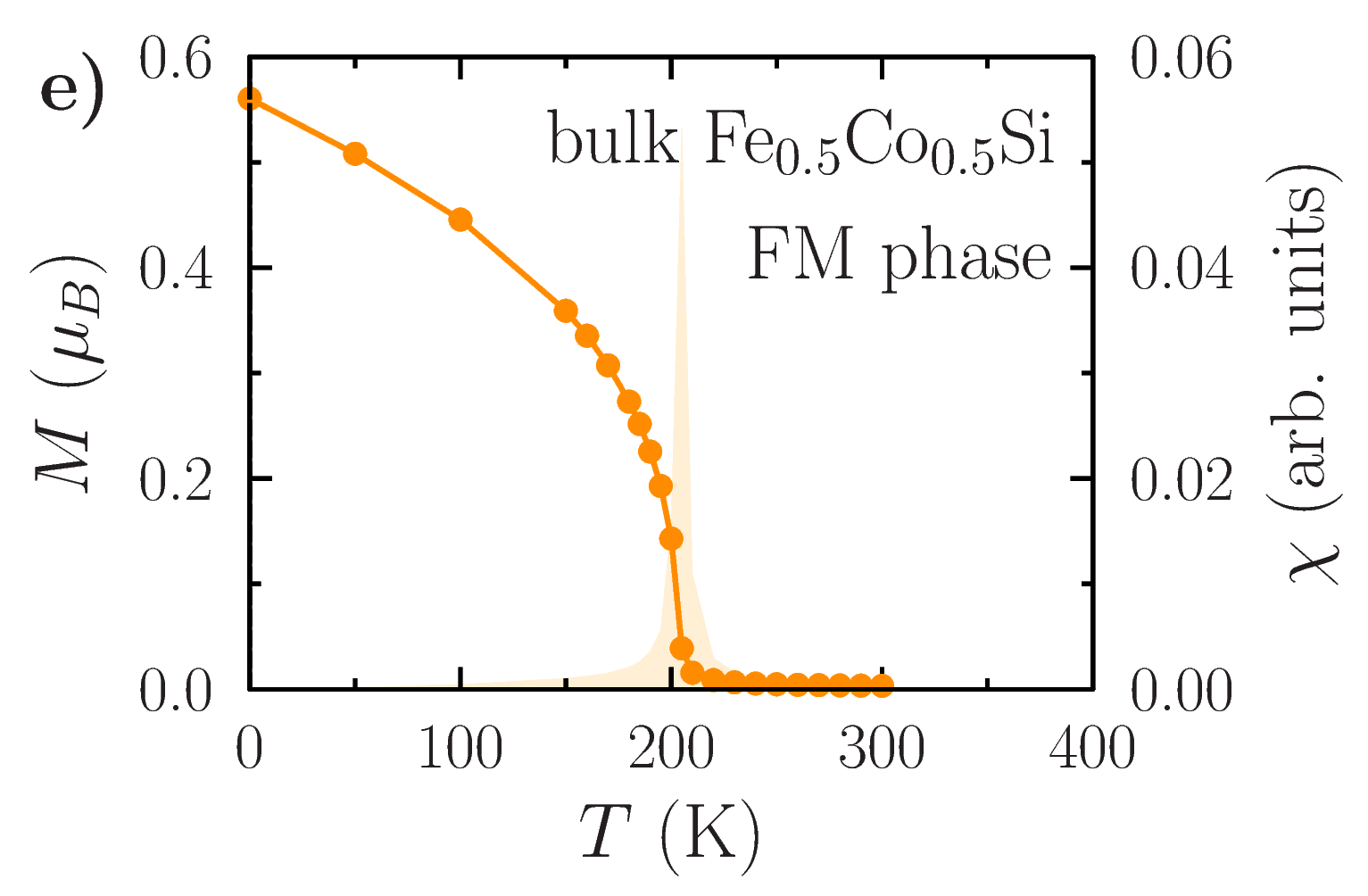}
\includegraphics[width=0.49\columnwidth]{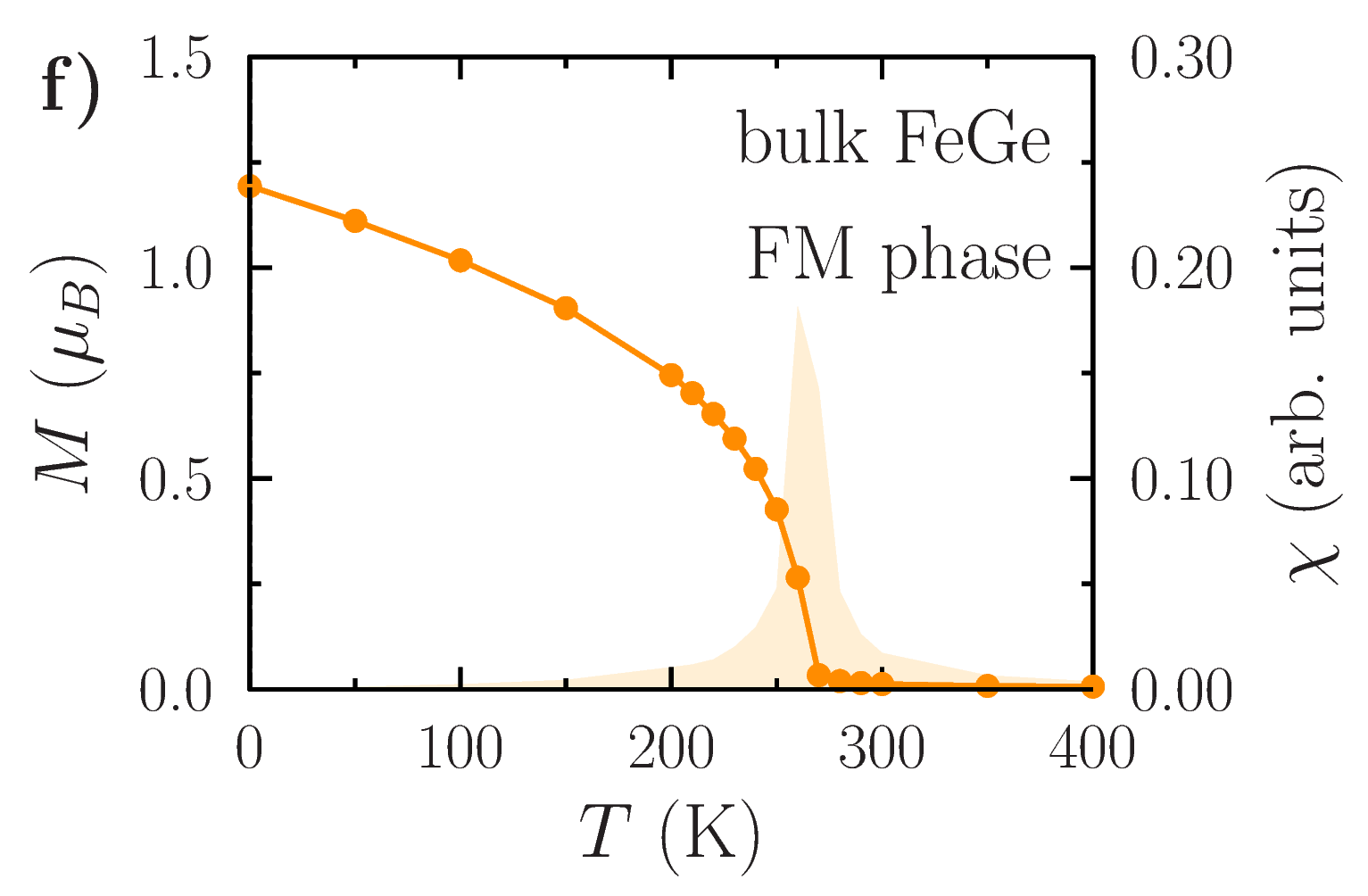}
}
\caption{Temperature-dependent total magnetization (per atom) of the ferromagnetic phase of the a) (FeSi)$_3$/(CoSi)$_3$, b) (FeSi)$_3$/(FeGe)$_3$ and d) (FeSi)$_3$/(FeGe)$_1$ superlattices and c) antiferromagnetic (FeSi)$_4$/(CoSi)$_1$ superlattice as well as bulk e) Fe$_{0.5}$Co$_{0.5}$Si and f) FeGe compounds obtained from the Monte Carlo simulations with \textit{ab initio} magnetic interactions.}
\label{f:M_vs_T}
\end{figure}

We performed Monte Carlo simulations using the calculated magnetic interactions for $(20\times20\times20)$ and $(30\times30\times30)$ simulation cells of different B20 multilayers and corresponding bulk compounds (selected results are shown in Fig.~\ref{f:M_vs_T}). The temperature-dependent magnetic order parameter indicate clearly the critical temperatures for long-range order for all studied systems. In most cases, the order parameter is just the total magnetization per atom in the unit cell, while for the (FeSi)$_4$/(CoSi)$_1$ superlattice the order parameter is the difference between the total magnetizations of neighboring interfaces, since the system is a layered magnet (Fig.~\ref{f:FeSi_CoSi_superlattices}b in the main text).

For FeSi/CoSi systems, we see that the magnetism is weaker than in the corresponding bulk compound. However, our calculations for bulk Fe$_{0.5}$Co$_{0.5}$Si overestimate significantly the Curie temperature ($\unit[200]{K}$ in Fig.~\ref{f:M_vs_T}e vs. $\unit[35]{K}$ in Ref.~\onlinecite{Onose2005}), which is a common problem of first-principles theory for some B20 compounds [\onlinecite{Collyer2008},\onlinecite{Grytsiuk2019}]. Despite this deficiency, our calculations for superlattices predict more realistic values for the critical temperature of long-range magnetic order (Fig.~\ref{f:M_vs_T}a,c).

For FeSi/FeGe systems, we predict stronger magnetism and higher critical temperature $T_\mathrm{C}$. For bulk FeGe, our prediction of $T_\mathrm{C} = \unit[260]{K}$ is in line with experimental result $\unit[280]{K}$ [\onlinecite{Spencer2018}]. The (FeSi)$_3$/(FeGe)$_1$ superlattice shows reduced moments near the interfaces and twice as low $T_\mathrm{C}=\unit[140]{K}$, while the (FeSi)$_3$/(FeGe)$_3$ superlattice shows almost the same strength of magnetism as bulk FeGe, indicating that the thickness of the FeGe layer of around $\unit[14]{\AA}$ is large enough to neglect the interface effects.

\section{Magnetic moments in $\mathrm{FeSi}/\mathrm{FeGe}$ superlattices}

\begin{figure}[h]
{\centering
\includegraphics[width=0.7\columnwidth]{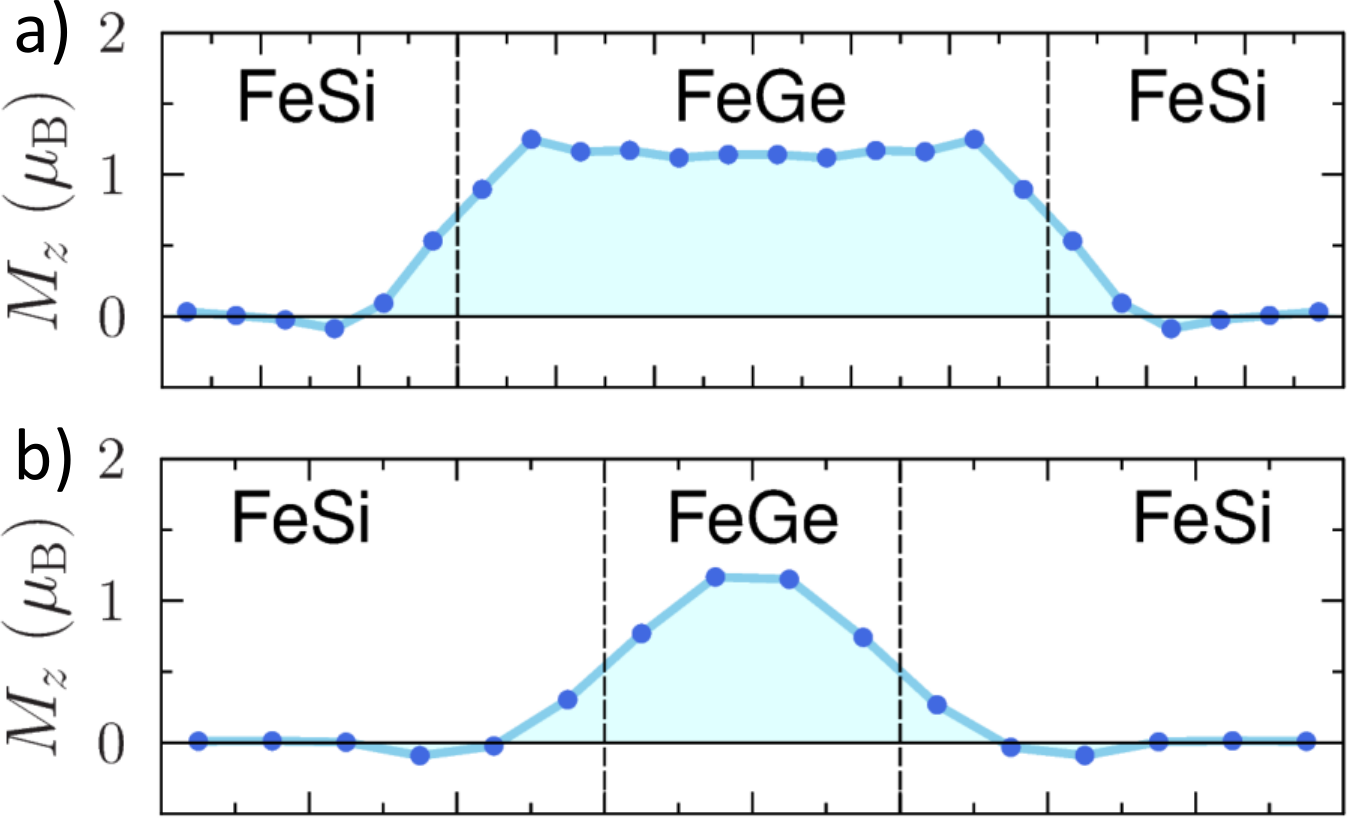}
}
\caption{Layer-resolved magnetic moments on the Fe sites in the a) (FeSi)$_3$/(FeGe)$_3$ and b) (FeSi)$_3$/(FeGe)$_1$ superlattices; the interfaces are marked by vertical dashed lines. Crystal structures are similar to those shown in Fig.~\ref{f:FeSi_CoSi_superlattices}.}
\label{f:FeSi_FeGe_superlattices}
\end{figure}

The distribution of magnetic moments in the [001]-oriented FeSi/FeGe superlattices is shown in Fig.~\ref{f:FeSi_FeGe_superlattices} where one can see the reduction of Fe moments in FeGe near the interface as well as the induction of smaller moments in the FeSi layer nearby. For both systems, the FeGe layers in the middle have moments similar to bulk, and the 3/3-superlattice clearly shows that the interface-induced changes in magnetization are restricted to just a few interfacial layers.

\section{Numerical aspects of calculating micromagnetic parameters}
\label{t:calc_micro}

From the first-principles magnetic interactions $J_{ij}$ and $\vec{D}_{ij}$ for hundreds of different pairs of spins at distance $\vec{R}_{ij}$ in (\ref{e:spin_model}), we determine the micromagnetic parameters spin stiffness $A$ and the $(3\times3)$ DM spiralization matrix $D_{\alpha\beta}$, according to the definitions
\begin{equation}
  A = \frac{1}{2}\sum_{j\neq i} J_{ij}R^2_{ij}\,e^{-\mu R_{ij}}, \hspace{2pt} D_{\alpha\beta} = \sum_{j\neq i} D_{ij}^\alpha R_{ij}^\beta \,e^{-\mu R_{ij}}.
  \label{e:micromagnetic_parameters}
\end{equation}

\begin{figure}[h]
\vspace{10pt}
{\centering
\includegraphics[width=0.6\textwidth]{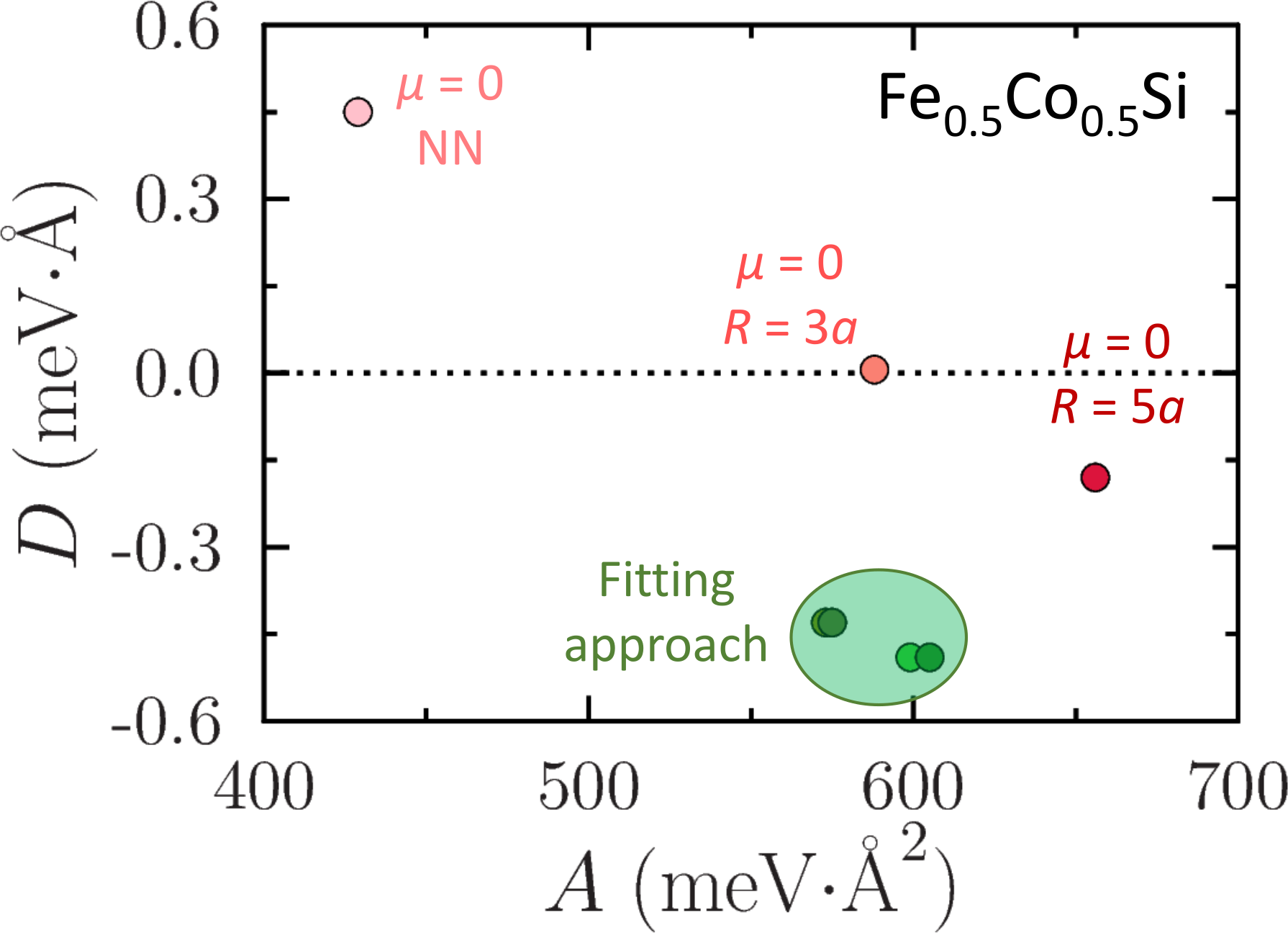}
}
\vspace{-5pt}
\caption{Spin stiffness ($x$-axis) and Dzyaloshinskii-Moriya micromagnetic parameter ($y$-axis) of bulk Fe$_{0.5}$Co$_{0.5}$Si calculated using two different approaches based on Eqns.~(\ref{e:micromagnetic_parameters}) in the main text: direct summation of atomistic interactions ($\mu = 0$) and extrapolation to $\mu\to0$ limit based on finite-$\mu$ data (details in the text). The later approach provides more reliable values of the micromagnetic parameters with less uncertainty (green-shaded area).}
\label{f:bulk_FeCoSi}
\end{figure}

Here, the exponential factor is introduced to improve the convergence of the sums and the limit $\mu\to 0$ is taken in the final step using 3$^\mathrm{rd}$-order polynomial extrapolation. Technical details of this calculation are discussed in SI and illustrated by Fig.~\ref{f:bulk_FeCoSi} as well as in Ref.~[\onlinecite{Borisov2023}] and in the SI of Ref.~\onlinecite{Borisov2022}.

Due to the long-range character of magnetic interactions in metallic systems, the sums over different atomic neighbors in Eqns.~\ref{e:micromagnetic_parameters} converge slowly with respect to the real-space cutoff, even when distances $R_{ij}$ between interacting spins up to 5 lattice parameters are considered. More than 20 years ago a solution to this problem was suggested [\onlinecite{Pajda2001}], which we also use in this work, by introducing an exponential decay factor $e^{-\mu R_{ij}}$ with positive parameter $\mu$. For sufficiently large $\mu$, the sums (\ref{e:micromagnetic_parameters}) start to converge to some well-defined values which become functions $A(\mu)$ (spin stiffnes) and $D(\mu)$ (DM matrix) of parameter $\mu$. We extrapolate these functions to the limit $\mu\to 0$ using a $3^\mathrm{rd}$-order polynomial, which gives reliable fits.

In Fig.~\ref{f:bulk_FeCoSi}, the results of this procedure are shown on the example of bulk Fe$_{1-x}$Co$_x$Si system for $x=0.50$. Based on the calculated atomistic magnetic interactions, we calculate the spin stiffness $A$ and bulk DM parameter $D$ following different methods:
\begin{itemize}
    \item direct summation with only nearest-neighbor interactions ($\mu=0$, NN in Fig.~\ref{f:bulk_FeCoSi});
    \item direct summation with interactions up to a distance of 3 lattice parameters ($\mu=0$, $R=3a$ in Fig.~\ref{f:bulk_FeCoSi});
    \item direct summation with interactions up to a distance of 5 lattice parameters ($\mu=0$, $R=5a$ in Fig.~\ref{f:bulk_FeCoSi});
    \item Fitting approach discussed above and originally suggested in Ref.~\onlinecite{Pajda2001} (shaded region in Fig.~\ref{f:bulk_FeCoSi} with extrapolation results for different ranges of $\mu$-values).
\end{itemize}

As we see clearly from these numerical findings, the direct summation of interactions using Eqns.~(\ref{e:micromagnetic_parameters}) without the exponential factor leads to non-convergent results and it is impossible to make reliable statements about the value of the micromagnetic parameters. In contrast, the fitting procedure from Ref.~\onlinecite{Pajda2001} allows to obtain more stable estimates of the $A$ and $D$ parameters. Moderate variations, shown by the shaded region in Fig.~\ref{f:bulk_FeCoSi}, are due to the different choice of the range of $\mu$-values that are used for extrapolating the micromagnetic parameters to the limit $\mu\to 0$.

\section{Micromagnetic parameters}

For all the studied superlattices, we have calculated the micromagnetic quantities spin stiffness $A$ and spiralization matrix $D_{\alpha\beta}$, which originate from the Heisenberg and Dzyaloshinskii-Moriya interactions, respectively. A few interesting observations can be made from these data (in addition to the discussion in the main text):
\begin{itemize}
    \item The properties of the 2/2-, 4/2- and 3/3-superlattices of FeSi/CoSi are quite similar (first 3 entries in Table I), which is not surprising, because magnetic moments are induced very close to the interface and the middle layers of FeSi and CoSi in both systems are virtually non-magnetic. For that reason, it should not make a lot of difference, if more layers are added to the superlattice.
    \item 3/1- and 4/1-superlattices are quite similar too, for the same reason, and, in terms of the $A/D$ ratio, they are also similar to the three systems discussed above. However, these systems have antiferromagnetically coupled interfaces and the structure of the DM matrix is distinctly different compared to the 2/2-, 4/2- and 3/3-superlattices discussed above.
    \item The [111]-oriented (FeSi)$_3$/(CoSi)$_3$ superlattice shows again a different symmetry of the DM matrix and larger $A/D$ ratio, suggesting a smaller relative magnitude of the DMI compared to the Heisenberg interaction, in comparison to the [001]-oriented superlattices.
    \item All these superlattices show a considerably stronger DM interaction compared to the bulk Fe$_{0.5}$Co$_{0.5}$Si compound, suggesting that nanoscale interfaces of B20 compounds can produce larger spin-orbit-related effects.
    \item For the FeGe-based multilayers, the presence of nanoscale interfaces does not seem to be produce stronger DM interaction, compared to the bulk FeGe, even though the symmetry of the DM matrix is quite different. Moreover, the critical temperature of the 3/1-superlattice is dramatically reduced, which can be attributed to dimensionality effects (quasi-2D vs 3D) and reduction of Fe magnetic moments near the interfaces.
\end{itemize}

Overall, based on our calculations, the AFM (FeSi)$_4$/(CoSi)$_1$ superlattice seems to have the strongest DM interaction (relative to the Heisenberg exchange, $A/D = \unit[3.65]{nm}$) among all the systems that we studied. On the other hand, all the FeSi/CoSi superllatices show significantly enhanced DMI (see the last column in Table~I in the SI) compared to the bulk Fe$_{1-x}$Co$_x$Si compound, meaning that such B20 multilayers can be promising candidates for topological magnetism with more compact textures.

\setlength{\tabcolsep}{6pt}
\renewcommand{\arraystretch}{1.5}
\begin{table}
 \caption{Calculated micromagnetic parameters (spin stiffness $A$ and DM matrix $D_{\alpha\beta}$) for different B20 multilayers and the corresponding bulk compounds. Unless otherwise specified, all multilayers have [001]-oriented interfaces. For systems with FeGe, the 4-site model takes into account only the spins in the FeGe layers, while the 6-site model also includes the induced moments in the FeSi layers near the interfaces. For the (FeSi)$_3$/(FeGe)$_3$ superlattice, the micromagnetic parameters are divided by 3 (due to 3 unit cells of magnetic FeGe) to simplify the comparison with other systems. Bulk Fe$_{0.5}$Co$_{0.5}$Si is calculated in two ways, using the virtual crystal approximation or digital alloy model. The last column shows the ratio between the spin stiffness and magnitude of the DM interaction, estimated as the sum of all elements of the DM matrix. This ratio has a unit of length and characterizes the scale of magnetic textures in the system; smaller values imply more compact magnetic objects (spin spirals, skyrmions etc.) and indicate more promising candidate systems.}
\vspace{5pt}
  \centering
  \begin{tabular}{c|c|c|c}
    \hline
    System & $A$ (meV$\cdot\mathrm{\AA}^2$) & $D_{\alpha\beta}$ (meV$\cdot\mathrm{\AA}$) & $A/\sum\limits_{\alpha\beta} D_{\alpha\beta}$ (nm) \\
    \hline
    (FeSi)$_2$/(CoSi)$_2$ & 169 &
    $
      \left(
        \begin{array}{ccc}
       0.83 &  0.36 & -0.35 \\
      -0.11 & -0.78 & -0.47 \\
       0.89 &  0.59 &  0.05 \\
        \end{array}
      \right)
    $ & 3.81 \\
    (FeSi)$_4$/(CoSi)$_2$ & 169 &  
    $
      \left(
        \begin{array}{ccc}
       0.78 &  0.39 & -0.31 \\
      -0.02 & -0.55 & -0.41 \\
       0.82 &  0.64 &  0.05 \\
        \end{array}
      \right)
    $ & 4.26 \\
    (FeSi)$_3$/(CoSi)$_3$ & 202 &  
    $
      \left(
        \begin{array}{ccc}
      0.86 &  0.37 & -0.27 \\
      0.02 & -0.69 & -0.36 \\
      1.08 &  0.65 & -0.19 \\
        \end{array}
      \right)
    $ & 4.50 \\\hline
    (FeSi)$_3$/(CoSi)$_1$ & 283 &  
    $   
      \left(
        \begin{array}{ccc}
       1.49 &  0.0  &  0.0  \\
       0.0  & -1.26 & -1.89 \\
       0.0  &  0.88 &  0.73 \\
        \end{array}
      \right)
    $ & 4.53 \\
    (FeSi)$_4$/(CoSi)$_1$ & 229 &
    $
      \left(
        \begin{array}{ccc}
      2.26 &  0.0  &  0.0  \\
      0.0  & -0.50 & -2.22 \\
      0.0  &  1.16 &  0.13 \\
        \end{array}
      \right)
    $ & 3.65 \\\hline
    (FeSi)$_3$/(CoSi)$_3$ [111] & 671 &  
    $
      \left(
        \begin{array}{ccc}
        -2.34  &  0.56  &  0.0  \\
        -0.55  & -2.35  &  0.0  \\
         0.0   &  0.0   & -0.55 \\
        \end{array}
      \right)
    $ & 10.6 \\\hline
    Bulk Fe$_{0.5}$Co$_{0.5}$Si (VCA) & 575 &  
    $
      \left(
        \begin{array}{ccc}
      -0.43 &  0.0  &  0.0  \\
       0.0  & -0.43 &  0.0  \\
       0.0  &  0.0  & -0.43 \\
        \end{array}
      \right)
    $ & 44.6 \\
    Bulk Fe$_{0.5}$Co$_{0.5}$Si (digital alloy) & 756 &  
    $
      \left(
        \begin{array}{ccc}
          -0.622 & -0.358 & -0.157 \\
          -0.157 & -0.622 & -0.358 \\
          -0.358 & -0.157 & -0.622 \\
        \end{array}
      \right)
    $ & 22.2 \\\hline
    (FeSi)$_3$/(FeGe)$_3$ & 639 &
    $
      \left(
        \begin{array}{ccc}
        0.32 &  0.00  &  0.0  \\
        0.0  & -0.89  & -0.02 \\
        0.0  & -0.42  & -2.61 \\
        \end{array}
      \right)
    $ & 15.0 \\[5pt]
    (FeSi)$_3$/(FeGe)$_1$ (4-site model) & 627 &  
    $
      \left(
        \begin{array}{ccc}
       0.07 &  0.11 &  0.12 \\
      -0.20 & -1.13 &  2.70\\
       0.03 & -1.25 & -0.74 \\
        \end{array}
      \right)
    $ & 9.9 \\ 
    (FeSi)$_3$/(FeGe)$_1$ (6-site model) & 723 &  
    $
      \left(
        \begin{array}{ccc}
       0.08 & -0.13 & -0.07 \\
      -0.05 & -1.85 &  0.80\\
       0.01 &  1.30 & -1.16 \\
        \end{array}
      \right)
    $ & 13.2 \\\hline
    Bulk FeGe & 1139 &  
    $
      \left(
        \begin{array}{ccc}
      -2.71 &  0.0  &  0.0  \\
       0.0  & -2.71 &  0.0  \\
       0.0  &  0.0  & -2.71 \\
        \end{array}
      \right)
    $ & 14.0 \\
    \hline
  \end{tabular}
  \label{t:micromagnetic_parameters}
\end{table}

\section{Effective micromagnetic field}
\label{t:micro_field}

The effective field is determined by the micromagnetic functional derived from the spin model (details in Ref.~[\onlinecite{Poluektov2018},\onlinecite{Borisov2023}]). For the general form of the DM matrix $D_{\alpha\beta}$, the effective field is defined as follows:
\begin{equation}
    B^\alpha_\mathrm{eff} = -\varepsilon_{\alpha\gamma\delta} D_{\gamma\beta}\nabla_\beta m_\delta,
    \label{e:effective_field}
\end{equation}

where $\varepsilon_{\alpha\gamma\beta}$ is the totally antisymmetric Levi-Civita symbol and the summation runs over repeated indices. Due to the quasi-2D character of the B20 systems that we consider in this work, one can, in principle, neglect the $z$-dependence of the magnetization, so all terms proportional to $\partial \vec{m}/\partial z$ can be omitted from Eqn.~\ref{e:effective_field}, which is equivalent to neglecting the last column of the DM matrix $D_{\alpha\beta}$ ($\beta = z$).

\medskip
\bibliographystyle{prb-titles}
\bibliography{main}

\begin{thebibliography}{10}
\providecommand{\bibAnnoteFile}[1]{%
  \IfFileExists{#1}{\begin{quotation}\noindent\textsc{Key:} #1\\
  \textsc{Annotation:}\ \input{#1}\end{quotation}}{}}
\providecommand{\bibAnnote}[2]{%
  \begin{quotation}\noindent\textsc{Key:} #1\\
  \textsc{Annotation:}\ #2\end{quotation}}
\providecommand{\bibinfo}[2]{#2}

\bibitem{Wood2000}
\bibinfo{author}{R.~Wood}, \bibinfo{journal}{IEEE Transactions on Magnetics}
  \textbf{\bibinfo{volume}{36}}, \bibinfo{pages}{36} (\bibinfo{year}{2000}).
\bibAnnoteFile{Wood2000}

\bibitem{Fert2013}
\bibinfo{author}{A.~Fert}, \bibinfo{author}{V.~Cros}, and
  \bibinfo{author}{J.~Sampaio}, \bibinfo{journal}{Nature Nanotech}
  \textbf{\bibinfo{volume}{8}}, \bibinfo{pages}{152–156}
  (\bibinfo{year}{2013}).
\bibAnnoteFile{Fert2013}

\bibitem{Jena2020}
\bibinfo{author}{J.~Jena}, \bibinfo{author}{B.~Göbel},
  \bibinfo{author}{T.~Ma}, \bibinfo{author}{V.~Kumar},
  \bibinfo{author}{R.~Saha}, \bibinfo{author}{I.~Mertig},
  \bibinfo{author}{C.~Felser}, and \bibinfo{author}{S.~S.~P. Parkin},
  \bibinfo{journal}{Nat. Commun.} \textbf{\bibinfo{volume}{11}},
  \bibinfo{pages}{1115} (\bibinfo{year}{2020}).
\bibAnnoteFile{Jena2020}

\bibitem{Huang2017}
\bibinfo{author}{Y.~Huang}, \bibinfo{author}{W.~Kang},
  \bibinfo{author}{X.~Zhang}, \bibinfo{author}{Y.~Zhou}, and
  \bibinfo{author}{W.~Zhao}, \bibinfo{journal}{Nanotechnology}
  \textbf{\bibinfo{volume}{28}}, \bibinfo{pages}{08LT02}
  (\bibinfo{year}{2017}).
\bibAnnoteFile{Huang2017}

\bibitem{Bourianoff2018}
\bibinfo{author}{G.~Bourianoff}, \bibinfo{author}{D.~Pinna},
  \bibinfo{author}{M.~Sitte}, and \bibinfo{author}{K.~Everschor-Sitte},
  \bibinfo{journal}{AIP Advances} \textbf{\bibinfo{volume}{8}},
  \bibinfo{pages}{055602} (\bibinfo{year}{2018}).
\bibAnnoteFile{Bourianoff2018}

\bibitem{Pinna2018}
\bibinfo{author}{D.~Pinna}, \bibinfo{author}{F.~Abreu~Araujo},
  \bibinfo{author}{J.-V. Kim}, \bibinfo{author}{V.~Cros},
  \bibinfo{author}{D.~Querlioz}, \bibinfo{author}{P.~Bessiere},
  \bibinfo{author}{J.~Droulez}, and \bibinfo{author}{J.~Grollier},
  \bibinfo{journal}{Phys. Rev. Appl.} \textbf{\bibinfo{volume}{9}},
  \bibinfo{pages}{064018} (\bibinfo{year}{2018}).
\bibAnnoteFile{Pinna2018}

\bibitem{Goebel2021}
\bibinfo{author}{B.~Göbel}, \bibinfo{author}{I.~Mertig}, and
  \bibinfo{author}{O.~A. Tretiakov}, \bibinfo{journal}{Physics Reports}
  \textbf{\bibinfo{volume}{895}}, \bibinfo{pages}{1} (\bibinfo{year}{2021}).
\bibAnnoteFile{Goebel2021}

\bibitem{Muehlbauer2009}
\bibinfo{author}{S.~M\"uhlbauer}, \bibinfo{author}{B.~Binz},
  \bibinfo{author}{F.~Jonietz}, \bibinfo{author}{C.~Pfleiderer},
  \bibinfo{author}{A.~Rosch}, \bibinfo{author}{A.~Neubauer},
  \bibinfo{author}{R.~Georgii}, and \bibinfo{author}{P.~B\"oni},
  \bibinfo{journal}{Science} \textbf{\bibinfo{volume}{323}},
  \bibinfo{pages}{915} (\bibinfo{year}{2009}).
\bibAnnoteFile{Muehlbauer2009}

\bibitem{Kezsmarki2015}
\bibinfo{author}{I.~Kézsmárki}, \bibinfo{author}{S.~Bordács},
  \bibinfo{author}{P.~Milde}, \bibinfo{author}{E.~Neuber},
  \bibinfo{author}{L.~M. Eng}, \bibinfo{author}{J.~S. White},
  \bibinfo{author}{H.~M. Rønnow}, \bibinfo{author}{C.~D. Dewhurst},
  \bibinfo{author}{M.~Mochizuki}, \bibinfo{author}{K.~Yanai},
  \bibinfo{author}{H.~Nakamura}, \bibinfo{author}{D.~Ehlers},
  \bibinfo{author}{V.~Tsurkan}, and \bibinfo{author}{A.~Loidl},
  \bibinfo{journal}{Nature Mater.} \textbf{\bibinfo{volume}{14}},
  \bibinfo{pages}{1116} (\bibinfo{year}{2015}).
\bibAnnoteFile{Kezsmarki2015}

\bibitem{Kanazawa2017}
\bibinfo{author}{N.~Kanazawa}, \bibinfo{author}{S.~Seki}, and
  \bibinfo{author}{Y.~Tokura}, \bibinfo{journal}{Advanced Materials}
  \textbf{\bibinfo{volume}{29}}, \bibinfo{pages}{1603227}
  (\bibinfo{year}{2017}).
\bibAnnoteFile{Kanazawa2017}

\bibitem{Soumyanarayanan2017}
\bibinfo{author}{A.~Soumyanarayanan}, \bibinfo{author}{M.~Raju},
  \bibinfo{author}{A.~L.~G. Oyarce}, \bibinfo{author}{A.~K.~C. Tan},
  \bibinfo{author}{M.-Y. Im}, \bibinfo{author}{A.~P. Petrović},
  \bibinfo{author}{P.~Ho}, \bibinfo{author}{K.~H. Khoo},
  \bibinfo{author}{M.~Tran}, \bibinfo{author}{C.~K. Gan},
  \bibinfo{author}{F.~Ernult}, and \bibinfo{author}{C.~Panagopoulos},
  \bibinfo{journal}{Nature Materials} \textbf{\bibinfo{volume}{16}},
  \bibinfo{pages}{898–} (\bibinfo{year}{2017}).
\bibAnnoteFile{Soumyanarayanan2017}

\bibitem{Heinze2011}
\bibinfo{author}{S.~Heinze}, \bibinfo{author}{K.~von Bergmann},
  \bibinfo{author}{M.~Menzel}, \bibinfo{author}{J.~Brede},
  \bibinfo{author}{A.~Kubetzka}, \bibinfo{author}{R.~Wiesendanger},
  \bibinfo{author}{G.~Bihlmayer}, and \bibinfo{author}{S.~Blügel},
  \bibinfo{journal}{Nature Physics} \textbf{\bibinfo{volume}{7}},
  \bibinfo{pages}{713–718} (\bibinfo{year}{2011}).
\bibAnnoteFile{Heinze2011}

\bibitem{Romming2013}
\bibinfo{author}{N.~Romming}, \bibinfo{author}{C.~Hanneken},
  \bibinfo{author}{M.~Menzel}, \bibinfo{author}{J.~E. Bickel},
  \bibinfo{author}{B.~Wolter}, \bibinfo{author}{K.~von Bergmann},
  \bibinfo{author}{A.~Kubetzka}, and \bibinfo{author}{R.~Wiesendanger},
  \bibinfo{journal}{Science} \textbf{\bibinfo{volume}{341}},
  \bibinfo{pages}{636} (\bibinfo{year}{2013}).
\bibAnnoteFile{Romming2013}

\bibitem{Bogdanov2001}
\bibinfo{author}{A.~N. Bogdanov} and \bibinfo{author}{U.~K. R\"o\ss{}ler},
  \bibinfo{journal}{Phys. Rev. Lett.} \textbf{\bibinfo{volume}{87}},
  \bibinfo{pages}{037203} (\bibinfo{year}{2001}).
\bibAnnoteFile{Bogdanov2001}

\bibitem{Li2014}
\bibinfo{author}{X.~Li}, \bibinfo{author}{W.~V. Liu}, and
  \bibinfo{author}{L.~Balents}, \bibinfo{journal}{Phys. Rev. Lett.}
  \textbf{\bibinfo{volume}{112}}, \bibinfo{pages}{067202}
  (\bibinfo{year}{2014}).
\bibAnnoteFile{Li2014}

\bibitem{Nandy2016}
\bibinfo{author}{A.~K. Nandy}, \bibinfo{author}{N.~S. Kiselev}, and
  \bibinfo{author}{S.~Bl\"ugel}, \bibinfo{journal}{Phys. Rev. Lett.}
  \textbf{\bibinfo{volume}{116}}, \bibinfo{pages}{177202}
  (\bibinfo{year}{2016}).
\bibAnnoteFile{Nandy2016}

\bibitem{Butler2001}
\bibinfo{author}{W.~H. Butler}, \bibinfo{author}{X.-G. Zhang},
  \bibinfo{author}{T.~C. Schulthess}, and \bibinfo{author}{J.~M. MacLaren},
  \bibinfo{journal}{Phys. Rev. B} \textbf{\bibinfo{volume}{63}},
  \bibinfo{pages}{054416} (\bibinfo{year}{2001}).
\bibAnnoteFile{Butler2001}

\bibitem{Bowen2001}
\bibinfo{author}{M.~Bowen}, \bibinfo{author}{V.~Cros},
  \bibinfo{author}{F.~Petroff}, \bibinfo{author}{A.~Fert},
  \bibinfo{author}{C.~Martinez~Boubeta}, \bibinfo{author}{J.~L. Costa-Krämer},
  \bibinfo{author}{J.~V. Anguita}, \bibinfo{author}{A.~Cebollada},
  \bibinfo{author}{F.~Briones}, \bibinfo{author}{J.~M. de~Teresa},
  \bibinfo{author}{L.~Morellón}, \bibinfo{author}{M.~R. Ibarra},
  \bibinfo{author}{F.~Güell}, \bibinfo{author}{F.~Peiró}, and
  \bibinfo{author}{A.~Cornet}, \bibinfo{journal}{Applied Physics Letters}
  \textbf{\bibinfo{volume}{79}}, \bibinfo{pages}{1655} (\bibinfo{year}{2001}).
\bibAnnoteFile{Bowen2001}

\bibitem{Jepsen1980}
\bibinfo{author}{O.~Jepsen}, \bibinfo{author}{J.~Madsen}, and
  \bibinfo{author}{O.~Andersen}, \bibinfo{journal}{Journal of Magnetism and
  Magnetic Materials} \textbf{\bibinfo{volume}{15-18}}, \bibinfo{pages}{867}
  (\bibinfo{year}{1980}).
\bibAnnoteFile{Jepsen1980}

\bibitem{Eriksson1991}
\bibinfo{author}{O.~Eriksson}, \bibinfo{author}{G.~Fernando},
  \bibinfo{author}{R.~Albers}, and \bibinfo{author}{A.~Boring},
  \bibinfo{journal}{Solid State Communications} \textbf{\bibinfo{volume}{78}},
  \bibinfo{pages}{801} (\bibinfo{year}{1991}).
\bibAnnoteFile{Eriksson1991}

\bibitem{Lau2002}
\bibinfo{author}{J.~T. Lau}, \bibinfo{author}{A.~Föhlisch},
  \bibinfo{author}{M.~Martins}, \bibinfo{author}{R.~Nietubyc},
  \bibinfo{author}{M.~Reif}, and \bibinfo{author}{W.~Wurth},
  \bibinfo{journal}{New Journal of Physics} \textbf{\bibinfo{volume}{4}},
  \bibinfo{pages}{98} (\bibinfo{year}{2002}).
\bibAnnoteFile{Lau2002}

\bibitem{Tischer1995}
\bibinfo{author}{M.~Tischer}, \bibinfo{author}{O.~Hjortstam},
  \bibinfo{author}{D.~Arvanitis}, \bibinfo{author}{J.~H. Dunn},
  \bibinfo{author}{F.~May}, \bibinfo{author}{K.~Baberschke},
  \bibinfo{author}{J.~Trygg}, \bibinfo{author}{J.~M. Wills},
  \bibinfo{author}{B.~Johansson}, and \bibinfo{author}{O.~Eriksson},
  \bibinfo{journal}{Phys. Rev. Lett.} \textbf{\bibinfo{volume}{75}},
  \bibinfo{pages}{1602} (\bibinfo{year}{1995}).
\bibAnnoteFile{Tischer1995}

\bibitem{Pfleiderer2001}
\bibinfo{author}{C.~Pfleiderer}, \bibinfo{author}{M.~Uhlarz},
  \bibinfo{author}{S.~M. Hayden}, \bibinfo{author}{R.~Vollmer},
  \bibinfo{author}{H.~v.~Löhneysen}, \bibinfo{author}{N.~R. Bernhoeft}, and
  \bibinfo{author}{G.~G. Lonzarich}, \bibinfo{journal}{Nature}
  \textbf{\bibinfo{volume}{412}}, \bibinfo{pages}{58–61}
  (\bibinfo{year}{2001}).
\bibAnnoteFile{Pfleiderer2001}

\bibitem{Aeppli1992}
\bibinfo{author}{G.~Aeppli} and \bibinfo{author}{Z.~Fisk},
  \bibinfo{journal}{Comments Condens. Matter Phys.}
  \textbf{\bibinfo{volume}{16}}, \bibinfo{pages}{155} (\bibinfo{year}{1992}).
\bibAnnoteFile{Aeppli1992}

\bibitem{Wernick1972}
\bibinfo{author}{J.~H. Wernick}, \bibinfo{author}{G.~K. Wertheim}, and
  \bibinfo{author}{R.~C. Sherwood}, \bibinfo{journal}{Mater. Res. Bull.}
  \textbf{\bibinfo{volume}{7}}, \bibinfo{pages}{1431} (\bibinfo{year}{1972}).
\bibAnnoteFile{Wernick1972}

\bibitem{Yu2010}
\bibinfo{author}{X.~Z. Yu}, \bibinfo{author}{Y.~Onose},
  \bibinfo{author}{N.~Kanazawa}, \bibinfo{author}{J.~H. Park},
  \bibinfo{author}{J.~H. Han}, \bibinfo{author}{Y.~Matsui},
  \bibinfo{author}{N.~Nagaosa}, and \bibinfo{author}{Y.~Tokura},
  \bibinfo{journal}{Nature} \textbf{\bibinfo{volume}{465}},
  \bibinfo{pages}{901–904} (\bibinfo{year}{2010}).
\bibAnnoteFile{Yu2010}

\bibitem{Yu2011}
\bibinfo{author}{X.~Z. Yu}, \bibinfo{author}{N.~Kanazawa},
  \bibinfo{author}{Y.~Onose}, \bibinfo{author}{K.~Kimoto},
  \bibinfo{author}{W.~Z. Zhang}, \bibinfo{author}{S.~Ishiwata},
  \bibinfo{author}{Y.~Matsui}, and \bibinfo{author}{Y.~Tokura},
  \bibinfo{journal}{Nature Materials} \textbf{\bibinfo{volume}{10}},
  \bibinfo{pages}{106} (\bibinfo{year}{2011}).
\bibAnnoteFile{Yu2011}

\bibitem{Borisov2022}
\bibinfo{author}{V.~Borisov}, \bibinfo{author}{Q.~Xu},
  \bibinfo{author}{N.~Ntallis}, \bibinfo{author}{R.~Clulow},
  \bibinfo{author}{V.~Shtender}, \bibinfo{author}{J.~Cedervall},
  \bibinfo{author}{M.~Sahlberg}, \bibinfo{author}{K.~T. Wikfeldt},
  \bibinfo{author}{D.~Thonig}, \bibinfo{author}{M.~Pereiro},
  \bibinfo{author}{A.~Bergman}, \bibinfo{author}{A.~Delin}, and
  \bibinfo{author}{O.~Eriksson}, \bibinfo{journal}{Phys. Rev. Materials}
  \textbf{\bibinfo{volume}{6}}, \bibinfo{pages}{084401} (\bibinfo{year}{2022}).
\bibAnnoteFile{Borisov2022}

\bibitem{Borisov2023}
\bibinfo{author}{V.~Borisov}, \bibinfo{author}{N.~Salehi},
  \bibinfo{author}{M.~Pereiro}, \bibinfo{author}{A.~Delin}, and
  \bibinfo{author}{O.~Eriksson}, \bibinfo{journal}{arXiv:2307.05733}
  (\bibinfo{year}{2023}).
\bibAnnoteFile{Borisov2023}

\bibitem{Hohenberg1964}
\bibinfo{author}{P.~Hohenberg} and \bibinfo{author}{W.~Kohn},
  \bibinfo{journal}{Phys. Rev.} \textbf{\bibinfo{volume}{136}},
  \bibinfo{pages}{B864} (\bibinfo{year}{1964}).
\bibAnnoteFile{Hohenberg1964}

\bibitem{Wills1987}
\bibinfo{author}{J.~M. Wills} and \bibinfo{author}{B.~R. Cooper},
  \bibinfo{journal}{Phys. Rev. B} \textbf{\bibinfo{volume}{36}},
  \bibinfo{pages}{3809} (\bibinfo{year}{1987}).
\bibAnnoteFile{Wills1987}

\bibitem{Wills2010}
\bibinfo{author}{J.~Wills~\textit{et al.}},
  \emph{\bibinfo{title}{Full-Potential Electronic Structure Method}}, volume
  \bibinfo{volume}{167}, \bibinfo{publisher}{Springer-Verlag Berlin Heidelberg}
  (\bibinfo{year}{2010}).
\bibAnnoteFile{Wills2010}

\bibitem{PBE1996}
\bibinfo{author}{J.~P. Perdew}, \bibinfo{author}{K.~Burke}, and
  \bibinfo{author}{M.~Ernzerhof}, \bibinfo{journal}{Phys. Rev. Lett.}
  \textbf{\bibinfo{volume}{77}}, \bibinfo{pages}{3865} (\bibinfo{year}{1996}).
\bibAnnoteFile{PBE1996}

\bibitem{ASE2002}
\bibinfo{author}{S.~Bahn} and \bibinfo{author}{K.~Jacobsen},
  \bibinfo{journal}{Computing in Science and Engineering}
  \textbf{\bibinfo{volume}{4}}, \bibinfo{pages}{56} (\bibinfo{year}{2002}).
\bibAnnoteFile{ASE2002}

\bibitem{ASE2017}
\bibinfo{author}{A.~H. Larsen}, \bibinfo{author}{J.~J. Mortensen},
  \bibinfo{author}{J.~Blomqvist}, \bibinfo{author}{I.~E. Castelli},
  \bibinfo{author}{R.~Christensen}, \bibinfo{author}{M.~Dułak},
  \bibinfo{author}{J.~Friis}, \bibinfo{author}{M.~N. Groves},
  \bibinfo{author}{B.~Hammer}, \bibinfo{author}{C.~Hargus},
  \bibinfo{author}{E.~D. Hermes}, \bibinfo{author}{P.~C. Jennings},
  \bibinfo{author}{P.~B. Jensen}, \bibinfo{author}{J.~Kermode},
  \bibinfo{author}{J.~R. Kitchin}, \bibinfo{author}{E.~L. Kolsbjerg},
  \bibinfo{author}{J.~Kubal}, \bibinfo{author}{K.~Kaasbjerg},
  \bibinfo{author}{S.~Lysgaard}, \bibinfo{author}{J.~B. Maronsson},
  \bibinfo{author}{T.~Maxson}, \bibinfo{author}{T.~Olsen},
  \bibinfo{author}{L.~Pastewka}, \bibinfo{author}{A.~Peterson},
  \bibinfo{author}{C.~Rostgaard}, \bibinfo{author}{J.~Schiøtz},
  \bibinfo{author}{O.~Schütt}, \bibinfo{author}{M.~Strange},
  \bibinfo{author}{K.~S. Thygesen}, \bibinfo{author}{T.~Vegge},
  \bibinfo{author}{L.~Vilhelmsen}, \bibinfo{author}{M.~Walter},
  \bibinfo{author}{Z.~Zeng}, and \bibinfo{author}{K.~W. Jacobsen},
  \bibinfo{journal}{Journal of Physics: Condensed Matter}
  \textbf{\bibinfo{volume}{29}}, \bibinfo{pages}{273002}
  (\bibinfo{year}{2017}).
\bibAnnoteFile{ASE2017}

\bibitem{Kresse1996}
\bibinfo{author}{G.~Kresse} and \bibinfo{author}{J.~Furthm\"uller},
  \bibinfo{journal}{Phys. Rev. B} \textbf{\bibinfo{volume}{54}},
  \bibinfo{pages}{11169} (\bibinfo{year}{1996}).
\bibAnnoteFile{Kresse1996}

\bibitem{LKAG1987}
\bibinfo{author}{A.~I. Liechtenstein}, \bibinfo{author}{M.~I. Katsnelson},
  \bibinfo{author}{V.~P. Antropov}, and \bibinfo{author}{V.~A. Gubanov},
  \bibinfo{journal}{J. Magn. Magn. Mater.} \textbf{\bibinfo{volume}{67}},
  \bibinfo{pages}{65 } (\bibinfo{year}{1987}).
\bibAnnoteFile{LKAG1987}

\bibitem{Szilva2023}
\bibinfo{author}{A.~Szilva}, \bibinfo{author}{Y.~Kvashnin},
  \bibinfo{author}{E.~A. Stepanov}, \bibinfo{author}{L.~Nordstr\"om},
  \bibinfo{author}{O.~Eriksson}, \bibinfo{author}{A.~I. Lichtenstein}, and
  \bibinfo{author}{M.~I. Katsnelson}, \bibinfo{journal}{Rev. Mod. Phys.}
  \textbf{\bibinfo{volume}{95}}, \bibinfo{pages}{035004}
  (\bibinfo{year}{2023}).
\bibAnnoteFile{Szilva2023}

\bibitem{uppasd}
\bibinfo{author}{B.~Skubic}, \bibinfo{author}{J.~Hellsvik},
  \bibinfo{author}{L.~Nordström}, and \bibinfo{author}{O.~Eriksson},
  \bibinfo{journal}{Journal of Physics: Condensed Matter}
  \textbf{\bibinfo{volume}{20}}, \bibinfo{pages}{315203}
  (\bibinfo{year}{2008}).
\bibAnnoteFile{uppasd}

\bibitem{Eriksson2017}
\bibinfo{author}{O.~Eriksson}, \bibinfo{author}{A.~Bergman},
  \bibinfo{author}{L.~Bergqvist}, and \bibinfo{author}{J.~Hellsvik},
  \emph{\bibinfo{title}{Atomistic Spin Dynamics: Foundations and
  Applications}}, \bibinfo{publisher}{Oxford University Press, Oxford, UK}
  (\bibinfo{year}{2017}).
\bibAnnoteFile{Eriksson2017}

\bibitem{Landau1935}
\bibinfo{author}{L.~Landau} and \bibinfo{author}{E.~Lifshitz},
  \bibinfo{journal}{Phys. Z. Sowjetunion} \textbf{\bibinfo{volume}{8}},
  \bibinfo{pages}{153} (\bibinfo{year}{1935}).
\bibAnnoteFile{Landau1935}

\bibitem{Gilbert2004}
\bibinfo{author}{T.~Gilbert}, \bibinfo{journal}{IEEE Trans. Mag.}
  \textbf{\bibinfo{volume}{40}}, \bibinfo{pages}{3443} (\bibinfo{year}{2004}).
\bibAnnoteFile{Gilbert2004}

\bibitem{Kim2020}
\bibinfo{author}{J.-V. Kim} and \bibinfo{author}{J.~Mulkers},
  \bibinfo{journal}{IOP SciNotes} \textbf{\bibinfo{volume}{1}},
  \bibinfo{pages}{025211} (\bibinfo{year}{2020}).
\bibAnnoteFile{Kim2020}

\bibitem{Lebech1989}
\bibinfo{author}{B.~Lebech}, \bibinfo{author}{J.~Bernhard}, and
  \bibinfo{author}{T.~Freltoft}, \bibinfo{journal}{Journal of Physics:
  Condensed Matter} \textbf{\bibinfo{volume}{1}}, \bibinfo{pages}{6105}
  (\bibinfo{year}{1989}).
\bibAnnoteFile{Lebech1989}

\bibitem{Uchida2008}
\bibinfo{author}{M.~Uchida}, \bibinfo{author}{N.~Nagaosa},
  \bibinfo{author}{J.~P. He}, \bibinfo{author}{Y.~Kaneko},
  \bibinfo{author}{S.~Iguchi}, \bibinfo{author}{Y.~Matsui}, and
  \bibinfo{author}{Y.~Tokura}, \bibinfo{journal}{Phys. Rev. B}
  \textbf{\bibinfo{volume}{77}}, \bibinfo{pages}{184402}
  (\bibinfo{year}{2008}).
\bibAnnoteFile{Uchida2008}

\bibitem{Zhang2016}
\bibinfo{author}{X.~Zhang}, \bibinfo{author}{Y.~Zhou}, and
  \bibinfo{author}{M.~Ezawa}, \bibinfo{journal}{Nature Communications}
  \textbf{\bibinfo{volume}{7}}, \bibinfo{pages}{10293} (\bibinfo{year}{2016}).
\bibAnnoteFile{Zhang2016}

\bibitem{vesta}
\bibinfo{author}{K.~Momma} and \bibinfo{author}{F.~Izumi},
  \bibinfo{journal}{Journal of Applied Crystallography}
  \textbf{\bibinfo{volume}{44}}, \bibinfo{pages}{1272} (\bibinfo{year}{2011}).
\bibAnnoteFile{vesta}

\bibitem{paraview}
\bibinfo{journal}{ParaView -- Open-source, multi-platform data analysis and
  visualization application (https://www.paraview.org/)} .
\bibAnnoteFile{paraview}

\bibitem{Onose2005}
\bibinfo{author}{Y.~Onose}, \bibinfo{author}{N.~Takeshita},
  \bibinfo{author}{C.~Terakura}, \bibinfo{author}{H.~Takagi}, and
  \bibinfo{author}{Y.~Tokura}, \bibinfo{journal}{Phys. Rev. B}
  \textbf{\bibinfo{volume}{72}}, \bibinfo{pages}{224431}
  (\bibinfo{year}{2005}).
\bibAnnoteFile{Onose2005}

\bibitem{Collyer2008}
\bibinfo{author}{R.~Collyer} and \bibinfo{author}{D.~Browne},
  \bibinfo{journal}{Physica B: Condensed Matter}
  \textbf{\bibinfo{volume}{403}}, \bibinfo{pages}{1420} (\bibinfo{year}{2008}).
\bibAnnoteFile{Collyer2008}

\bibitem{Grytsiuk2019}
\bibinfo{author}{S.~Grytsiuk}, \bibinfo{author}{M.~Hoffmann},
  \bibinfo{author}{J.-P. Hanke}, \bibinfo{author}{P.~Mavropoulos},
  \bibinfo{author}{Y.~Mokrousov}, \bibinfo{author}{G.~Bihlmayer}, and
  \bibinfo{author}{S.~Bl\"ugel}, \bibinfo{journal}{Phys. Rev. B}
  \textbf{\bibinfo{volume}{100}}, \bibinfo{pages}{214406}
  (\bibinfo{year}{2019}).
\bibAnnoteFile{Grytsiuk2019}

\bibitem{Spencer2018}
\bibinfo{author}{C.~S. Spencer}, \bibinfo{author}{J.~Gayles},
  \bibinfo{author}{N.~A. Porter}, \bibinfo{author}{S.~Sugimoto},
  \bibinfo{author}{Z.~Aslam}, \bibinfo{author}{C.~J. Kinane},
  \bibinfo{author}{T.~R. Charlton}, \bibinfo{author}{F.~Freimuth},
  \bibinfo{author}{S.~Chadov}, \bibinfo{author}{S.~Langridge},
  \bibinfo{author}{J.~Sinova}, \bibinfo{author}{C.~Felser},
  \bibinfo{author}{S.~Bl\"ugel}, \bibinfo{author}{Y.~Mokrousov}, and
  \bibinfo{author}{C.~H. Marrows}, \bibinfo{journal}{Phys. Rev. B}
  \textbf{\bibinfo{volume}{97}}, \bibinfo{pages}{214406}
  (\bibinfo{year}{2018}).
\bibAnnoteFile{Spencer2018}

\bibitem{Pajda2001}
\bibinfo{author}{M.~Pajda}, \bibinfo{author}{J.~Kudrnovsk\'y},
  \bibinfo{author}{I.~Turek}, \bibinfo{author}{V.~Drchal}, and
  \bibinfo{author}{P.~Bruno}, \bibinfo{journal}{Phys. Rev. B}
  \textbf{\bibinfo{volume}{64}}, \bibinfo{pages}{174402}
  (\bibinfo{year}{2001}).
\bibAnnoteFile{Pajda2001}

\bibitem{Poluektov2018}
\bibinfo{author}{M.~Poluektov}, \bibinfo{author}{O.~Eriksson}, and
  \bibinfo{author}{G.~Kreiss}, \bibinfo{journal}{Computer Methods in Applied
  Mechanics and Engineering} \textbf{\bibinfo{volume}{329}},
  \bibinfo{pages}{219} (\bibinfo{year}{2018}).
\bibAnnoteFile{Poluektov2018}

\end{thebibliography}

\end{document}